\documentclass[11pt]{article}

\usepackage{authblk}  % have affiliations with article
\usepackage{graphicx} % Include figure files
\usepackage{epsfig}
\usepackage{dcolumn}  % Align table columns on decimal point
\usepackage{bm}       % bold math
\usepackage{amsmath}
\usepackage{amssymb}
\usepackage{url}

\usepackage{float}

\usepackage{multirow}
\usepackage{booktabs}

\newcommand{\newsection}{ \setcounter{equation}{0} \section}

\newcommand{\beq}{\begin{equation}} \newcommand{\eeq}{\end{equation}}
\newcommand{\bea}{\begin{eqnarray}} \newcommand{\eea}{\end{eqnarray}}

  \newcommand
{\Romannumeral}[1]{\uppercase\expandafter{\romannumeral#1}}

\newcommand{\be}{\begin{enumerate}} \newcommand{\ee}{\end{enumerate}}
\newcommand{\bi}{\begin{itemize}} \newcommand{\ei}{\end{itemize}}
\newcommand{\ba}{\begin{array}} \newcommand{\ea}{\end{array}}
\newcommand{\bc}{\begin{center}} \newcommand{\ec}{\end{center}}
\newcommand{\bt}{\begin{tabular}} \newcommand{\et}{\end{tabular}}

\def\lsim{\mathrel{\rlap{\lower4pt\hbox{\hskip1pt$\sim$}}
    \raise1pt\hbox{$<$}}}           % less than or approx. symbol
\def\gsim{\mathrel{\rlap{\lower4pt\hbox{\hskip1pt$\sim$}}
    \raise1pt\hbox{$>$}}}           % greater than or approx. symbol

\newcommand{\half}{\textstyle {1\over2} \displaystyle}    % One half
\newcommand{\Dslash}{{\hbox{D}\kern-0.6em\raise0.15ex\hbox{/}}} % D slash

\renewcommand{\et}{\eta}

\begin{document}
\thispagestyle{empty} % suppresses display 1st page's number
	
\setlength{\oddsidemargin}{0cm}
\setlength{\baselineskip}{7mm}

\begin{normalsize}\begin{flushright}
Oct. 2019
\end{flushright}\end{normalsize}

\begin{center}

\vspace{15pt}

{\Large \bf Gravitational Fluctuations as an Alternative to Inflation II. 
\newline 
CMB Angular Power Spectrum}

\vspace{30pt}

{\sl Herbert W. Hamber ${}^a$ \footnote{HHamber@uci.edu.}, 
Lu Heng Sunny Yu ${}^{a,b}$ \footnote{Lhyu1@uci.edu.}} 
\\
${}^a$ Department of Physics and Astronomy \\
University of California \\
Irvine, CA 92697-4575, USA
\\
${}^b$ Theory Division \\
SLAC National Accelerator Laboratory \\
Sand Hill Road \\
Menlo Park, CA 94309, USA
\\

\vspace{10pt}
\end{center}

%  ABSTRACT 

\begin{center} 
{\bf ABSTRACT } 
\end{center}

\noindent

Power spectra always play an important role in the theory of inflation.  
In particular, the ability to reproduce the galaxy matter power spectrum $ P(k) $ 
and the CMB temperature angular power spectrum $ C_l $'s to high accuracy is 
often considered a triumph of inflation.
In our previous work, we presented an alternative explanation for the matter power spectrum 
based on nonperturbative quantum field-theoretical methods applied to Einstein's gravity, 
instead of inflation models based on scalar fields.  
In this work, we review the basic concepts and provide further in-depth investigations.  
We first update the analysis with more recent data sets and error analysis, and
then extend our predictions to the CMB angular spectrum coefficients $ C_l $, 
which we did not consider previously. 
Then we investigate further the potential freedoms and uncertainties associated with 
the fundamental parameters that are part of this picture, and show how recent cosmological 
data provides significant constraints on these quantities.  
Overall, we find good general consistency between theory and data, even potentially 
favoring the gravitationally-motivated picture at the largest scales.  
We summarize our results by outlining how this picture can be tested in the near future 
with increasingly accurate astrophysical measurements.

% Keywords : Quantum Cosmology, Quantum Gravity, Inflationary Cosmology

% Preprints 2019, 2019100101 
% doi: 10.20944/preprints201910.0101.v1
% arXiv:1910.02990 [gr-qc]

\newpage

\section{Introduction}
\label{sec:intro}  

\vskip 10pt

In cosmology we know that the Universe is not perfectly homogeneous and isotropic 
but rather comprises of fluctuations, in matter density and in temperature, 
which are congregated and correlated in a rather specific manner.  
Detailed measurements of these fluctuations can be characterized by correlation 
functions and power spectra
% \cite{pee93,pee98,bah03,bau06,lon07,teg02,teg04,dur14,wan13,coi12}.  
[1-10].
The question of why these fluctuations are distributed the way they are is thus an 
important one in cosmology.
The conventional explanation of the shape of these power spectra is provided by inflation, 
which is based on the hypothesis of additional primordial scalar fields called inflatons 
\cite{gut81, lin82, alb82}.  
The shape of the power spectrum is thus derived from quantum fluctuation of these 
primordial inflaton fields, and the agreement of this prediction with observations 
to high accuracy has been widely regarded as a great triumph and confirmation for 
inflation \cite{lid00}.  

In our previous work \cite{hyu18}, we have offered an alternative explanation 
based on gravitational fluctuations alone without inflation, which to our 
knowledge is the first-of-its-kind.  
While the short-distance theory of quantum gravity may still be highly uncertain 
due to both the flexibility of higher-order operators consistent with general 
covariance and the lack of experimental results, the long-distance or infrared limit 
of the theory is however in principle well-defined and unique, governed largely 
by the concept of universality.  
Although this long-distance quantum theory of gravity still suffers from 
being perturbatively nonrenormalizable, well known field theory techniques have 
been extensively developed, applied and tested in many other fields of 
physics where perturbation theory fails, usually due to a non-trivial vacuum structure.  
As a result, it is thus conceivable that these nonperturbative methods may find use 
in deriving physical consequences of perturbatively non-renormalizable theories such 
as gravity.  
Previous efforts \cite{ham17, book} have shown that many such effects may manifest 
themselves and become important on very large cosmological scales.  
In particular, we found that much of the matter power spectrum can be derived 
and reproduced from Einstein gravity and standard $\Lambda$CDM cosmology alone, 
utilizing nonperturbative quantum field methods, without the need of 
additional scalar fields as advocated by inflation.  
We have shown that not only the predictions agree quite well with recent data, 
but also that additional quantum effects predict subtle deviations from the 
classical picture, which may become testable in the near future.

In this paper, we further investigate consequences from the above picture.  
Most importantly, we translate our quantum gravity prediction of the 
matter wavenumber power spectrum $P(k)$ to a prediction for the angular 
temperature power spectrum coefficients $C_l$'s.   
The angular temperature power spectrum, which expresses the power 
as coefficients of spherical harmonics (instead of plane waves, like the matter 
power spectrum) serves as a more direct comparison with observation, 
since after all the CMB measurements are performed over the sky.  
Unsurprisingly, we again find a general agreement between the observations 
and the gravitationally motivated prediction.  
Furthermore, additional quantum gravitational effects are expected to affect the 
low-$l$ regime of the $C_l$ spectrum.  
As discussed in \cite{hyu18}, new quantum effects become significant 
when the separation-distances $r$ become comparable to a characteristic 
vacuum condensate scale of gravity $ \xi $, which is expected to be 
extremely large ($\sim 5300 \, \rm{Mpc}$), affecting very small $l$'s in 
angular harmonic space.  
The additional quantum effects include the infrared (IR) regulator effects 
from the gravitational vacuum condensate and the renormalization group (RG) 
running of Newton's constant $G$.  
Here these effects on the angular spectrum are studied, and we show the occurrence 
of a dip in power in the low-$l$ regime.  
Furthermore, we argue that this may potentially be the cause for the well-known 
$l=2$ anomaly in the $C_l$ spectrum.  
Although it is impossible to make conclusive statements so far due to the large cosmic 
variance in that region, one might hope that there may be incremental improvements 
in systematic and experimental uncertainties in the near future, or that certain 
statistical likelihood-arguments for this effect can be made.

In addition, a number of updates will be presented here.  
In particular, new data sets of the matter power spectrum $P\left(k\right)$ have been 
released by the Planck collaboration \cite{planck18}, very shortly after the publication 
of our first results and predictions.  
It is interesting here to study the consistency and improvements, if any, of the data.  
We find that the refined analyses and results not only remain consistent with our 
theoretical result, but a new point was published in \cite{planck18}, 
which seems to suggest a downwards dip on the spectrum at low-$k$, as a 
running Newton's constant due to quantum gravity would suggest.  
Although the error bars on the points are too large to make conclusive 
statements, this remains an interesting development to study.

It is also possible to utilize these latest cosmological observations to constrain 
the theoretical values of the microscopic parameters, and thus further shed 
insight into the underlying theory.  
A handful of parameters in the theory are investigated, including the universal 
critical scaling exponent $\nu$, the coefficient for the amplitude of quantum 
effects $c_0$, and the characteristic nonperturbative correlation length scale $\xi$.  
Our analyses show that while the latter ($c_0$ and $\xi$) may only be 
constrained up to an order of magnitude, the data actually puts rather 
stringent constraints on the universal scaling exponent $\nu $.

Finally, we present a study of the effect of the RG running of Newton's constant $G$ 
derived from a recent alternative analytical nonperturbative 
approach - the Hartree-Fock (HF) approximation \cite{hyu19hf} - on the power spectra.  
We find that these results predict the same general indication of a decrease in the power 
spectrum amplitude at scales close to the characteristic scale $r \sim \xi$, 
a behavior which is consistent with the predictions from other nonperturbative 
methods such as the Regge-Wheeler lattice formulation and the $ 2 + \epsilon $ dimensional 
expansion results.  
However the HF approximation seems to eventually predict an upturn in 
power at extreme low-$k$ regime and diverge from the other methods, 
which presumably indicates the limit of validity for this particular approximation method.

The paper is organized as follows.  
Section 2 serves to outline the key points and results from the quantum 
theory of gravity and the resulting non-trivial scaling dimensions.  
Section 3 summarizes the main results in deriving the power spectrum from quantum gravity.  
This section also includes updated plots with the latest results from 
experiments, and updated error bars.
Section 4 and 5 investigate the possibility of constraining the theoretical 
scaling parameters from cosmology.  
Section 6 relates the predictions on the matter power spectrum $P(k)$ to the 
angular temperature spectrum $C_l $'s, and the effects of an RG running of 
Newton's constant $G$ on the $C_l $'s.  
Section 7 discusses, summarizes and contrasts the current quantum gravity motivated
picture with that of inflation,  in view of explaining the measured power spectra.  
The section concludes by outlining a number of future issues of interest to this study.

\vskip 20pt

\section{Nonperturbative Approach to Quantum Gravity}
\label{sec:introQG} 

\vskip 10pt

Many more details on the nonperturbative approach to quantum gravity used in this paper can be 
found in a number of earlier works \cite{ham17, book}, and many references therein.
The current section will therefore only serve to summarize the key points and main results which will become relevant for the subsequent discussion.

Quantum gravity, in essence the covariantly quantized theory of a massless spin-two particles, 
is in principle a unique theory, as shown by Feynman some time ago \cite{fey63, fey95}, 
much like Yang-Mills theory and QED are for massless spin-one particles.  
In the covariant Feynman path integral approach, only two key ingredients are needed to 
formulate the quantum theory - the gravitational action $ S \left[ g_{\mu\nu} \right] $ 
and the functional measure over metrics $ d \left[ g_{\mu\nu} \right] $, 
leading to the generating function  
\begin{equation}
Z \left[ g_{\mu\nu} \right] =
\int  d \left[ g_{\mu\nu} \right] e^{ \frac{i}{\hbar} S \left[ g_{\mu\nu} \right] } \;\; ,
\end{equation}
where all physical observables could in principle be derived from.  
For gravity the action is given by the Einstein-Hilbert term augmented by a cosmological constant  
\begin{equation}
S \left[ g_{\mu\nu} \right] = 
\frac{1}{16 \pi G} \int d^4 x  \sqrt{g}  \left( R - 2 \lambda \right) \;\; ,
\end{equation}
where $R$ is the Ricci scalar, $g$ being the determinant of the metric $ g_{\mu\nu} (x) $, 
$G$ Newton's constant, and $ \lambda $ the scaled cosmological constant
(where a lower case is used here, as opposed to the more popular upper case in cosmology, 
so as not to confuse 
it with the ultraviolet-cutoff in quantum field theories that is commonly associated with 
$\Lambda$).  
The other key ingredient is the functional measure for the metric field, which in 
the case of gravity describes an integration over all four metrics,
with weighting determined by the celebrated DeWitt form \cite{dew62}.  

There are two important subtleties worth noting here.  
Firstly, in principle, additional higher derivative terms that are consistent with general 
covariance could be allowed in the action, but nevertheless will only affect the physics 
at very short distances and will not be necessary here for studying large-distance cosmological effects.  
Secondly, as in most cases that the Feynman path integral can be written down, from 
non-relativistic quantum mechanics to field theories, the formal definition of integrals 
requires the introduction of a lattice, in order to properly account for, the known fact 
that quantum paths are nowhere differentiable.
It is therefore a remarkable aspect that, at least in principle, the theory, in a 
nonperturbative context, does not seem to require any additional extraneous 
ingredients, besides the standard ones mentioned above, to properly define a 
quantum theory of gravity.

At the same time, gravity does present some rather difficult and fundamentally inherent 
challenges, such as the perturbatively nonrenormalizable nature due to a badly 
divergent series in Newton's constant $G$, the intensive computational power
 required from being a highly nonlinear theory, the conformal instability which makes the 
Euclidean path integral potentially divergent, and further genuinely gravitational 
technical complications arising from the fact that physical distances between 
spacetime points,  which depend on the metric which is a quantum entity, fluctuate.  

Although these hurdles will ultimately need to be addressed in a complete 
and satisfactory way, a comprehensive account is of course far beyond the scope of this paper.  
However, regarding the perturbatively nonrenormalizable nature, some of the 
most interesting phenomena in physics often stem from non-analytic behavior in the 
coupling constant and the existence of nontrivial quantum condensates, which 
are hidden and impossible to probe within perturbation theory.  
It is therefore possible that challenges encountered in the case of gravity are more 
likely the result of inadequate calculational treatments, and not necessarily a 
reflection of some fundamentally insurmountable problem with the theory itself.  
Here, we shall take this as a motivation to utilize the plethora of well-established 
nonperturbative methods to deal with other quantum field theories where 
perturbation theory fails, and attempt to derive sensible physical predictions 
that can hopefully be tested against observations.  
More detailed accounts on the various issues associated with the theory of quantum 
gravity can be found for example in \cite{ham17, book}, and references therein.

For the present discussion, we will mention that the nonperturbative treatment of 
quantum gravity via Wilson's $ 2 + \epsilon $ double expansion (in $G$ and the dimension) 
and the Regge-Wheeler lattice path integral formulation \cite{lesh84} both reveal the existence of a 
new quantum phase, involving a nontrivial gravitational vacuum condensate \cite{ham17}.  
Along with this comes a nonperturbative characteristic correlation length scale, $\xi$, 
and a new set of non-trivial scaling exponents, as is common for well-studied 
perturbatively non-renormalizable theories 
$\nu$ 
% \cite{wil72,par73,par76,par81,itz91,car96,zin02,bre10}.  
[24-31].
Together, these two parameters characterize quantum corrections to physical observables 
such as the long-distance behavior of invariant correlation functions, as well as the 
renormalization group (RG) running of Newton's constant $G$, which in coordinate
space leads to a covariant
$G (\Box)$ with $\Box = g^{\mu\nu} \nabla_\mu \nabla_\nu $ \cite{book}.
In particular, in can be shown \cite{ham17, cor94} that for $r<\xi$, the correlation 
functions of the Ricci scalar curvatures over large geodesic separation 
$r\equiv\left|x-y\right|$ scales as 
\begin{equation}
G_R(r) \; = \; \langle \; \delta R(x) \; \delta R(y) \; \rangle \sim \; \frac{1}{r^{2(d-1/\nu)}} \;\; ,
\label{eq:GRcorr_scaling}
\end{equation}
where $d$ here the dimension of spacetime.  
Furthermore, the RG running of Newton's constant can be expressed as
\begin{equation}
G(k) \; = \; G_0 
\left[ \, 1 + 2 \; c_0 \left( \frac{m^2}{k^2} \right)^\frac{1}{2\nu}
+ \mathcal{O} \left( \left( \frac{m^2}{k^2} \right)^\frac{1}{\nu} \right) \,
\right]
\label{eq:Grun1}
\end{equation}
where $2 \, c_0 \approx 16.04$ is a nonperturbative coefficient, which can be
computed from first principles using the Regge-Wheeler lattice formulation of quantum gravity
\cite{hw05, rei10, rei14,ham15}.  

Here we note the important role played by the quantum parameters $\nu$ and $\xi$.  
The appearance of a gravitational condensate is viewed as analogous to the 
(equally nonperturbative) gluon and chiral condensates known to describe 
the physical vacuum of QCD, so that the genuinely nonperturbative 
scale $ \xi $ is in many ways analogous to the scaling violation parameter 
$ \Lambda_{\bar{MS}} $ of QCD.
Such a scale cannot be calculated from first principles, but should instead 
be linked with other length scales in the theory, such as the cosmological 
constant scale $\sqrt{1/\lambda}$.  
The combination that is most naturally identified with $\xi$ would be 
$ \xi \sim \sqrt{\rm{3/}\lambda} \simeq 5300 \, \rm{Mpc} $ \cite{loops,ham17}.
The other key quantity, the universal scaling dimension $\nu$, 
can be evaluated via a number of methods, many of which are summarized in 
% \cite{ham15,wei79,gas78,eps,larged,htw12,reu98,lit04,reu14,fal15,per16,gie15}.
[36,38-48].
Multiple avenues point to an indication of $\nu^{-1} \simeq 3$, which here will serve as 
a good working value for this parameter; a simple geometric argument suggests 
$\nu = 1 / (d-1) $ for spacetime dimension $d \ge 4$ \cite{book}.

It should be noted that the nonperturbative scale $\xi$ should also act as an infrared (IR) 
regulator, such that, like in other quantum field theories, expressions in 
the "infrared" (i.e. as $r \rightarrow \infty$, or equivalently $k \rightarrow 0$) 
should be augmented by
\begin{equation}
\frac{1}{k^2}
\rightarrow
\frac{1}{k^2+m^2}
\label{eq:IRreg}
\end{equation}
where $m = 1/\xi \simeq 2.8 \times 10^{-4} \, h \, \rm{Mpc}^{-1}$, 
expressed in the dimensionless Hubble constant $h\simeq 0.67$ for later convenience. 
Consequently, the augmented expression for the running of Newton's constant $G$ becomes
\begin{equation}
G(k) = G_0 \left[ \; 1 + 2 \; c_0 \left( \frac{m^2}{k^2+m^2} \right)^\frac{1}{2\nu}
+ \mathcal{O} \left(\left(\frac{m^2}{k^2+m^2}\right)^\frac{1}{\nu}\right) \; \right]
\;\; .
\label{eq:Grun2}
\end{equation}
The aim here is therefore to explore areas where these predictions can be put to a test.  
The cosmological power spectra, which are closely related to correlation functions, 
and thus take effects over large distances, provide a great testing ground for these 
quantum gravity effects.

\vskip 20pt

\section{Deriving the Matter Power Spectrum}
\label{sec:QGexpofPk}  

\vskip 10pt

The most straightforward relation to something testable is via the galaxy power spectrum.  
In this section, we provide a brief summary of the theory, as well as updated 
plots of observational results.  More details can be found in our previous paper \cite{hyu18}.  
A power spectrum is a correlation function in Fourier or wavenumber space.  
Thus, the galaxy power spectrum essentially quantifies how galaxies of various
separations are correlated \cite{pee93,pee98}.
More specifically, consider the distribution of galaxies described by a matter density field
$\rho(\mathbf{x},t)$. 
The overdensity, or fluctuation, above the average background density 
$\bar{\rho}$ can be quantified by the density contrast $\delta$, defined by
\begin{equation}
\delta (\mathbf{x},t) \equiv \frac{\delta\rho(\mathbf{x},t)}{\bar{\rho}(t)} 
= \frac{\rho(\mathbf{x},t) - \bar{\rho}}{\bar{\rho}(t)} \; \; .
\label{eq:delta_def}
\end{equation}
Correlations of such fluctuations between two points can be measured by the two-point 
matter density correlation function $G_\rho (r)$, defined as
\begin{equation}
G_\rho (r;t,t') 
\equiv 
\left\langle \, \delta(\mathbf{x},t) \,\, \delta(\mathbf{y},t') \, \right\rangle 
= \frac{1}{V} \int_{V} d^3\mathbf{z} \; \delta(\mathbf{x+z},t) \; \delta(\mathbf{y+z},t) \;\; ,
\label{eq:key}
\end{equation}
with $r=\left|\mathbf{x-y}\right|$.  
The above correlation function can also be expressed in Fourier-, or wavenumber-, space, 
$G_\rho (k;t,t') \equiv \left\langle \, \delta (k,t) \, \delta (-k,t') \, \right\rangle$, 
via a Fourier transform, as recalled below. 
For our analysis, it is useful to bring these measurements to a common time, 
say $t_0$, so that one can compare density fluctuations of different scales 
as they are measured and appear today.  
The resultant object $P(k)$ is referred to as the matter power spectrum,
\begin{equation}
P(k) \equiv 
(2\pi)^3 \langle \; \left| \delta (k,t_0) \right|^2 \; \rangle =
(2\pi)^3  F(t_0)^2   \langle \; \left| \Delta (k,t_0) \right|^2 \; \rangle \;\; ,
\label{eq:key1}
\end{equation}
where $\delta(k,t) \equiv F(t) \Delta(k,t_0)$.  
The factor $F(t)$ then simply follows the standard GR evolution formulas as governed 
by the Freidman-Robertson-Walker (FRW) metric.  
As a result, $P(k)$ can be related to, and extracted from, the real-space measurements 
via the inverse transform
\begin{equation}
\begin{aligned}
G_\rho (r;t,t') 
& =
\int \frac{d^3 k}{(2\pi)^3} \; G_\rho (k;t,t') \; 
e^{-i \mathbf{k} \cdot (\mathbf{x}-\mathbf{y})} \\
& = 
\frac{1}{2 \pi^2} \frac{F(t)F(t')}{F(t_0)^2} \; 
\int_{0}^{\infty} dk \; k^2 \; P(k) \; \frac{\sin{(kr)}}{kr}  \;\;  .
\end{aligned}
\label{eq:FT_PktoGrhor}
\end{equation}
It is often convenient to parameterize these correlators by a so-called 
scale-invariant spectrum, 
which includes an amplitude and a scaling index, conventionally written as 
\begin{equation}
P(k) = \frac{a_0}{k^s}  \;\;  ,
\label{eq:s_def}
\end{equation}
\begin{equation}
G_\rho (r;t_0,t_0 ) = \left( \frac{r_0}{r} \right) ^ \gamma \;\; .
\label{eq:gamma_def}
\end{equation}
It is then straightforward to relate the scaling indices using Eq.~(\ref{eq:FT_PktoGrhor}), 
giving $\gamma = 2 \, ( d - 1 / \nu ) = 2 $ and
\begin{equation}
s = (d-1) - \gamma = 3 - \gamma  = 1 \;\;  .
\label{eq:stogamma}
\end{equation} 
Note that $ G_\rho (r; t_0, t_0 ) $ is sometimes referred to as $ \xi (r) $ in the literature, 
but here we prefer to avoid confusion with the gravitational correlation 
length $\xi$ which will appear later.

Next, to relate the predictions from quantum gravity on the fluctuations of curvature to 
the fluctuations of the galaxy matter density, we make use of the Einstein field equations
\begin{equation}
R_{\mu\nu} \, - \, \half \, g_{\mu\nu} R + \lambda g_{\mu\nu} 
\, = \, 8 \pi G \, T_{\mu\nu} \;\; .
\label{eq:EFE}
\end{equation}
To a first approximation, by assuming galaxies follows a perfect pressureless fluid, 
the trace equation reads
\begin{equation}
R - 4 \lambda = - 8 \pi G \, T  \;\; .
\label{eq:EFEtrace}
\end{equation}
(For a perfect fluid the trace gives $T = 3 p - \rho$, and thus 
$ T \simeq - \rho $ for a non-relativistic fluid.)  
Since $\lambda$ is a constant, the variations, and hence correlations, are directly related as in
\begin{equation}
\langle \; \delta R \, \delta R \; \rangle \, = \, 
(8 \pi G)^2 \; \langle \; \delta \rho \, \delta \rho \; \rangle  \;\; .
\label{eq:GRtoGrho}
\end{equation}
As described in the previous section, quantum gravity predicts that the scalar curvature 
fluctuations $ G_R  $ over large distances scale as
\begin{equation}
G_R (r)  \equiv \langle \; \delta R (\mathbf{x}) \, \delta R (\mathbf{y}) \; \rangle
\, \sim \, \frac{1}{r^2}  \;\;  .
\label{eq:GRscaling}
\end{equation}
Using the relation in Eq. (\ref{eq:GRtoGrho}), the galaxy density fluctuations 
$ G_\rho  $ then follow a similar scaling relation
\begin{equation}
G_\rho  = \left( \frac{r_0}{r} \right)^2
\label{eq:Grhoscaling}
\end{equation}
as $r \rightarrow \infty$; or, using Eq.~(\ref{eq:stogamma}), equivalently,
\begin{equation}
P(k) = \frac{a_0}{k}
\label{eq:Pkscaling}
\end{equation}
as $k \rightarrow 0$ for the galaxy power spectrum.  
As a result, quantum gravity predicts an exponent $\gamma =2 \, (d-1/\nu) = 2$, 
or $s= d-1-\gamma = 1$ in $d=4$.  
The prediction of $\gamma =2$ or $s=1$ is expected to be valid in a so-called linear 
scaling regime, where galaxies can be treated as essentially pressureless point 
particles, and where long-range gravitational correlations are expected to be dominant.  
Quantitatively one observes that typical galaxy clusters have sizes around 
$r \sim 3-10$ Mpc (or equivalently $k \sim 2 \pi / r \sim 3.0-0.9 \, \rm{Mpc}^{-1} $).  
For separations below this scale, nonlinear dynamics is expected to dominate,
but beyond separations $r > 50-100$ Mpc ($k < 0.2-0.1 \, h \, \rm{Mpc}^{-1} $), 
any effects of 
such nonlinear dynamics should become unimportant, and long-range gravitational 
correlations are expected to dominate.  
It is over these large distances that one expects gravitationally induced scaling to 
take effect.

Nonlinear effects can be expected to modify the scaling in two ways.  
Firstly, including pressure to the Freidman equations induces fluctuations 
about the general scaling trend, known as baryonic acoustic oscillations (BAOs).  
Secondly, for distance scales below the size of galaxy clusters, nonlinear multi-body 
dynamics become important.  
Nevertheless, despite the computational complexity, such nonlinear dynamics basically 
follow Newtonian dynamics and are thus well-understood and well-studied in standard 
literatures such as \cite{pee93,wei08,dod03,ste04}.  
At these scales, neither quantum nor general-relativistic effects are expected to 
play a huge role.
% Figure 1
\begin{figure}
\begin{center}
\includegraphics[width=0.90\textwidth]{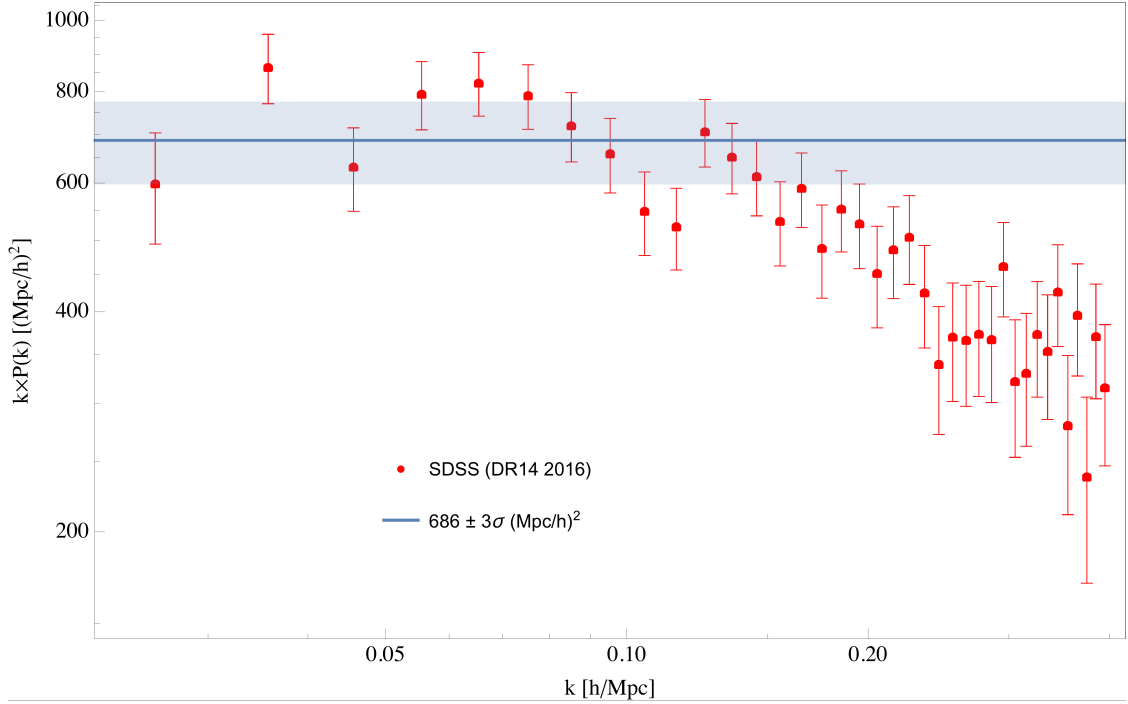}
\end{center}
\caption{
The observed galaxy power spectrum in $k \times P(k)$ versus wavenumber $ k $.  
The data points are taken from the Sloan Digital Sky Survey (SDSS) 
collaboration's 14th Data Release (DR14) \cite{gil18}.  
Quantum gravity predicts that in the linear regime ($ k \lsim 0.15 \, h \, \rm{Mpc}^{-1} $), 
as $ r \rightarrow \infty $ (or $ k \rightarrow 0 $), $ P(k) $ should approach a 
scale-invariant spectrum with $ \nu=1/3 $ (i.e. $ s=1 $), as in Eq. (\ref{eq:Pkscaling}).  
In other words, $ k \times P(k) $ should approach a constant.  
The solid line represents the asymptotic value of the $ s=1 $ spectrum, with a one-parameter 
fit for the amplitude in Eq. (\ref{eq:Pkscaling}) giving 
$ a_0 \simeq 686 \left( \rm{Mpc} / h \right)^2 $.  
The gray bands represent a $ 3\sigma $ ($\sim \pm 15$ \%) variance to the fit.  
It can be seen that, below $ k < 0.15 \, h \, Mpc^{-1} $, 
the data generally approach a constant of approximately 
$ a_0 \sim 686\ \left( \rm{Mpc} / h \right)^2 $, but beyond $ k > 0.15 \, h \, \rm{Mpc}^{-1} $ 
the data shows a transient region where the points deviate from the linear scaling, 
due to the relevant correlation function probing distances smaller than the linear scaling regime.
}
\label{fig:Pubplot1_sdssfitamp}
\end{figure} 
From the observational side, Fig. \ref{fig:Pubplot1_sdssfitamp} shows recent results 
of the power spectrum from the 14th data release (DR14) of the Sloan Digital Sky Survey (SDSS), 
a galaxy survey which charted over 1.5 million galaxies, covering over one-third of 
the celestial sphere \cite{gil18}, with separations roughly from 
$ k \simeq 0.02 \, h \, \rm{Mpc}^{-1}  $ ($ r \simeq 500 \, \rm{Mpc} $) to 
$ k \simeq 0.30 \, h \, \rm{Mpc}^{-1}  $ ($ r \simeq 30 \, \rm{Mpc} $).  
Notice as $k$ decreases ($r$ increases), the data appears to approach a constant 
on the $k \times P(k)$ vs. $k$ plot, which agrees with an $s=1$ scaling law
\begin{equation}
k \cdot P(k) = a_0 \;\; .
\label{eq:Pkscaling2}
\end{equation}
One then obtains a value of $ a_0 \approx 686 \pm 87 \, ( \rm{Mpc} / h )^2 $, 
which parameterizes the amplitude of the galaxy power spectrum.  
In particular, focusing on the linear scaling regime $ r \gsim 50 \rm{Mpc} $ 
($ k \lsim 0.15 \, h \, \rm{Mpc}^{-1} $), all of the corresponding 13-15 data points lie 
within a $ 3 \sigma $ ($ \sim 15\% $) variance of $ a_0 $.  
On the other hand, as expected, for separation distances smaller than 
$ 50 \, \rm{Mpc}  $ ($ k \gsim 0.17 \, h \, \rm{Mpc}^{-1} $), the spectrum deviates 
from the $ s=1 $ scaling law, giving rise to a transient behavior into the nonlinear regime.  
In addition, rough oscillations from BAOs can also be observed about the 
average value given by $ a_0 $.

One can further extend the above analysis by doing a phenomenological fit over 
the linear regime with two parameters, $a_0$ and $s$, 
using again the scale-invariant ansatz $ P(k)=a_0/k^s $.  
Such fit gives $s \simeq 1.09 \pm 0.08$, i.e. about $ 9.0 \, ( \, \pm \, 8.0) \% $ 
around the predicted $s=1$ value, and $a_0 = 540 \pm 115 \, ( \rm{Mpc} / h )^2$.  
This is a decent expectation given a first order prediction, neglecting BAOs and other 
dynamical effects superimposed on the linear scaling.  
Further analysis by applying the fit over the full range of observational data 
($ k=(0.02,0.30) \, h \, \rm{Mpc}^{-1} $) gives $s \simeq 1.31 \pm 0.04$, i.e. about 
$ 30 \% $ of the predicted $s=1$ value, and $a_0=280 \pm 24 \, ( \rm{Mpc} / h )^2$.  
The larger discrepancy in $s$ is also expected, given that the nonlinear regime is 
now included in the fit.  
Nevertheless, it is still overall consistent with an 
$s \sim 1$ trend, satisfying the general trend created and set by the gravitational correlations.
To even more accurately extrapolate the results to the nonlinear regime 
($ k > 0.15 \, h \, \rm{Mpc}^{-1} $), the full nonlinear dynamics has to be addressed and solved.  
In fact, we will see that the nonlinear solutions can be extrapolated to even 
larger scales ($ k < 0.02  \, h \, \rm{Mpc}^{-1} $) into a radiation dominated era of the Universe.  
This will be the topic of the next section.

It should be noted here that the amplitude $ a_0 $, just like the scaling 
dimension $\nu$ or $s$, is in principle calculable from the lattice treatment of 
quantum gravity, as discussed for example in \cite{ham17} and references therein.
Nevertheless, unlike the universal, scheme-independent, scaling dimension $s$, 
$a_0$ represents a non-universal quantity, and will therefore depend to some 
extend on the specific way the ultraviolet cutoff is implemented in the quantum theory.  
This fact is already well known from other lattice gauge theories such as lattice QCD.  
Therefore it seems more appropriate here to take this non-universal amplitude 
$a_0$ as an adjustable parameter, to be fitted and constrained by astrophysical 
observational data.  
Nevertheless, quantum gravity provides a direct prediction for the general 
$s=1$ scaling of galaxy correlations.

Notice that since galaxy scales from the SDSS survey are of the order 
$ 50-500 \, \rm{Mpc} $,
which is at least one to two orders of magnitude below 
$ \xi \simeq 5300 \, \rm{Mpc} $, 
the RG running of Newton's $G$ as governed by Eq.~(\ref{eq:Grun1}) is highly 
suppressed, and Newton's constant can be treated as a constant.  
Later on we will explore these additional effects as we turn to fluctuations on even 
larger scales in the next section.

\vskip 20pt

\section{Matter Power Spectrum in the Small-$k$ Regime}
\label{sec:Pksmallk} 

\vskip 10pt

It is possible to extrapolate the quantum gravity predictions in the linear regime 
of galaxies to both larger and smaller $k$ via a solution of the appropriate nonlinear 
evolution equations.  
The calculation is one that is rather complex algebraically and is often done via numerical 
programs such as CAMB.  
A semi-analytical treatment can be found in \cite{wei08}, which was adopted in our 
previous paper, and which will be used here as well.  
More details of the calculation can be found throughout our previous work \cite{hyu18}, 
and we will simply outline the key steps in this section, as well as present the latest 
observational results.

Already mentioned is the challenge of extending to smaller, nonlinear, scales 
($ k > 0.15 \, h \, \rm{Mpc}^{-1}  $).  
But a second challenge is to extend to larger distances ($ k < 0.02 \, h \, \rm{Mpc}^{-1}  $).  
Larger distances in the sky correspond to earlier epochs of the Universe, which 
eventually transits from a matter dominated one to a radiation dominated one, 
which takes place around $ k_{eq} \simeq 0.014 \, h \, \rm{Mpc}^{-1}  $ \cite{planck18}.  
For such distances ($ k < k_{eq} $) which correspond to a Universe that is constitute 
predominately out of radiation, it will be difficult to find fully formed galaxies.
Therefore the main source of observational data that involve such large 
separations comes from the observed cosmic microwave background (CMB).

However, the map of the CMB represents fluctuations in temperature, which 
are essentially fluctuations in radiation, not matter, density.
The quantum gravity prediction of $ G_R \sim 1/r^2 $ for the scaling of 
curvature fluctuations is in principle valid for all scales of separations 
(up to $ r < \xi = 5300 \, \rm{Mpc} $, or $ k > 1.8 \times 10^{-3} \, h \, \rm{Mpc}^{-1} $).  
However, whereas the scalar curvature correlation $ G_R $ is easily 
related to the matter density correlation $ G_\rho $ using the trace of the 
Einstein field equations in $ k > k_{eq} $ (where the Universe is matter dominated), 
it is not easy to relate to a radiation correlation $ G_{rad} $ for $ k < k_{eq} $ 
(where the Universe starts to become radiation dominated), since the trace of the 
energy-momentum tensor for radiation is zero.  
In this case the full Einstein field equations, not just their trace, is needed.

Both challenges can be circumvented via existing numerical calculations 
of the nonlinear cosmic evolution equations.  
For minimal confusion, we will strictly adhere to the notation in \cite{wei08} for 
the derivations and expressions for $ P(k) $, and later $ C_l $'s, 
unless otherwise stated.  
Their general solution for $ P(k) $ is given by the form
\begin{equation}
P(k) = C_0 \, 
\left( \mathcal{R}^0_k \right)^2 k^4 \, \left[ \mathcal{T}(\kappa) \right]^2  \;\;  ,
\label{eq:Pkfull}
\end{equation}
where $ C_0 \equiv 4 (2 \pi )^2 C^2 ( \Omega_\Lambda / \Omega_M ) / 25 \, 
\Omega_M^2 H_0^4 $, a prefactor of cosmological parameters.  
An initial condition is imposed on the factor $ \mathcal{R}_k^0 $, and 
the rest is the so-called transfer function $ \mathcal{T}(\kappa) $, 
a well-known result from standard cosmology literature \cite{whu97}.  
A semi-analytical formula for $ \mathcal{T}(\kappa) $ is given by
\begin{equation}
\mathcal{T} ( \kappa ) \simeq \frac{\ln[1+(0.124\kappa)^2]}{(0.124\kappa)^2} 
\left [  
{ {1+(1.257\kappa)^2+(0.4452\kappa)^4+(0.2197\kappa)^6} 
\over  
{1+(1.606\kappa)^2+(0.8568\kappa)^4+(0.3927\kappa)^6} }  
\right ]^{1/2} \; \; ,
\label{eq:Tk}
\end{equation}
in \cite{wei08}, with $ \kappa = \sqrt{2} \, k / k_{eq} $, and will be used 
here for simplicity.  
The remaining initial condition function $ \mathcal{R}_k^0 $ is usually 
parameterized in the form of a scale-invariant spectrum \cite{hz70} 
\begin{equation}
( \mathcal{R}_k^0 )^2 = N^2 \frac{1}{k^3} \left( \frac{k}{ k_{ \mathcal{R}} } \right)^{n_s - 1}
\label{eq:Rksq_param}
\end{equation}
which then forms the only ungoverned part of the power spectrum $ P(k) $.  
In other words, the full power spectrum, over a full scale of $k$, is 
now parameterized by only two theoretical parameters, $ N^2 $ and $ n_s $.  
The rest is then fully governed by classical physics and measured cosmological parameters.  
The quantity $ k_{ \mathcal{R} } $ is sometimes referred to as the ``pivot scale'', 
and here is simply a reference scale, conventionally taken to be 
$ k_{ \mathcal{R} } = 0.05 \, \rm{Mpc}^{-1}$.

\begin{figure}
\begin{center}
\includegraphics[width=0.90\textwidth]{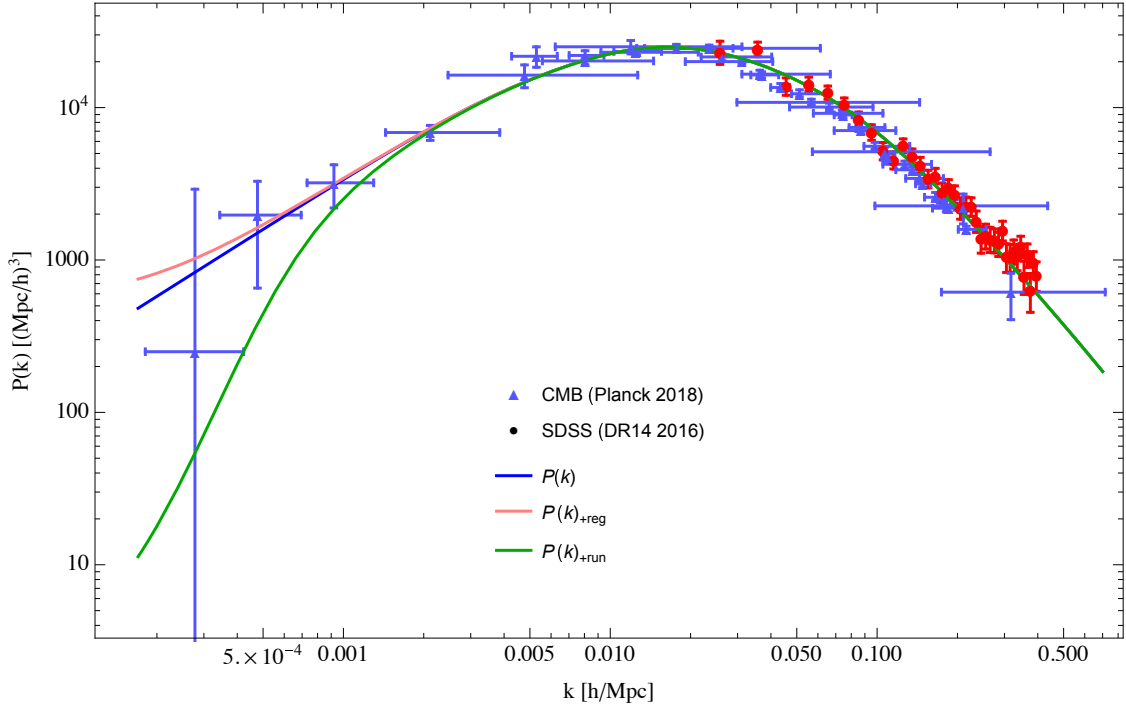}
\end{center}
\caption{
Matter power spectrum $ P(k) $ with various quantum effects included.  
The middle (blue) curve shows the full matter power spectrum function 
as predicted by quantum gravity, modulo the two following effects.  
The top (orange) curve $ P(k)_\text{reg} $ includes the effect of an infrared (IR) 
regulator $ \xi \sim 5300 \, \rm{Mpc} $.  
The bottom (green) curve $ P(k)_\text{run} $ includes both the effect of the IR regulator 
and a renormalization group (RG) running of Newton's $ G $, characterized by an amplitude 
of running $ 2 \, c_0 \sim 16.04 $, as given by numerical calculations from the 
lattice theory [see Eqs.~(\ref{eq:Grun1}) and (\ref{eq:Grun2})].  
The blue triangles show the Planck 2018 CMB data \cite{planck18}.  
}
\label{fig:Pubplot2_pkregrun}
\end{figure}

One can now normalize the spectrum from the galaxy regime (see \cite{hyu18} for details), 
and then the full power spectrum up to 
$ r < \xi = 5300 $ Mpc ($ k > 1.8 \times 10^{-3} \, h \, \rm{Mpc}^{-1} $) is fully determined.  
Fig. \ref{fig:Pubplot2_pkregrun} shows this prediction by the middle blue curve 
with the CMB data from the Planck satellite data (2018) \cite{planck18}, 
combined with the earlier galaxy data from SDSS (DR 14).  
As one can see, the middle blue curve is in good agreement with all the current 
CMB and Galaxy data points.  
This shows that applying quantum gravity results, together with standard cosmology 
derived from the Boltzmann transport equations and general relativity, provides a 
complete derivation of the power spectrum $ P(k) $.
% Figure 2
Now, at scales above $ r \sim \xi = 5300 $ Mpc (or $ k $ below 
$ m \sim 2.8 \times 10^{-4} \, h \, \rm{Mpc}^{-1} $), additional 
quantum effects are expected to be significant in the quantum theory of gravity.  
These effects can form potentially testable predictions for this picture which deviate 
from the current classical predictions.  Indeed, at small k, two additional genuinely 
quantum effects becomes important.  
Firstly, similar to QCD, an infrared (IR) regulator must be included to regulate 
expressions near vanishing $ k $.  
This is implemented, in close analogy to QCD, by promoting in various expressions
\begin{equation}
\frac{1}{k^2} \rightarrow \frac{1}{k^2+m^2}
\label{eq:IRreg2}
\end{equation}
where the scale $ m \sim 1/ \xi \sim 2.8 \times 10^{-4} \, h \, \rm{Mpc}^{-1} $.  
Secondly, the effects of renormalization group (RG) running of coupling 
constants will become significant at small $k$, which involves promoting 
Newton's constant $G$ to 
\begin{equation}
G \rightarrow G(k) = G_0 \left [ \;
1 + 2 c_0 \left( \frac{m^2}{k^2+m^2} \right)^{3/2} 
+ \mathcal{O} \left( \left( \frac{m^2}{k^2+m^2} \right)^{3} \right)
\right]  \;\; .
\label{eq:promoteGrun}
\end{equation}
Both effects are plotted in Fig. \ref{fig:Pubplot2_pkregrun}, with the top pink curve 
showing the first effect of an IR regulator, and the bottom green curve incorporating 
both effects together.  
Note that the combined effect is a dip below the classical (blue curve) results at 
scales of around $ k \sim m $.  
The deviation in these curves can potentially form a testable prediction of the 
quantum gravity picture, with increasingly precise results expected in the near future.

It is worth comparing the above results to the inflation-motivated picture.  
In essence, the inflation picture postulates a scalar field that is dominant in the 
early Universe (with some power-law potential), 
whose quantum fluctuations sets the scale of the fluctuations of the (Newtonian) 
gravitational potential $ \Phi $.  
The inflaton field then needs to be ``turned-off'', after some $ N_e $ number of $ e $-foldings, 
and the spectrum for gravity $ \Phi $ continues to govern the spectrum for matter, 
resulting in the observed matter spectrum today.  
It should be noted that the success of inflation has been largely attributed to its 
ability in explaining this power spectrum.  
Prior to inflation, the origin of this spectrum has long been a mystery.  
At the time, models such as a fractal Universe\cite{pee93} provided the best 
motivation for the Harrison-Zel'dovich spectrum of $ P(k) \sim 1/k $ in the galaxy regime.  
Nevertheless, the origin of this fractal behavior remained a mystery.  
With the invention of inflation, finally comes a quantum theory that is able to set the 
initial scale, and provides an explanation with a "primordial spectrum" that leads to 
$ ( \mathcal{R}_k^0 )^2 = N^2 k^{-3} (k / k_* )^{n_s-1} $ 
having a value $ n_s = 1 - 2 / N_e \approx 0.96 $ for basic scalar models assuming 
$N_e \approx 60$ $ e $-foldings.  
This prediction, which provided the first explanation for the shape of the power spectrum, 
has thus been presented as a triumph of inflation \cite{ste04, ste06}.

Here we have presented an alternate picture, whereby treating the 
gravitational field quantum mechanically one is able to extract the 
gravitational scalar curvature spectrum $ G_R $, which then directly governs the 
power spectrum for matter $ P(k) $.  
The scaling of the gravitational spectrum $ G_R $ is, whether evaluated 
numerically or estimated analytically, fully calculable from first principles of 
quantum field theory.  
Furthermore, the relation from the curvature spectrum to the matter spectrum 
$ P(k) $ is also in principle unambiguous, as presented in this work.

It is interesting to note that the scalar spectral index $n_s$ can also be 
extracted from this picture, which leads to some interesting comparison.  
The scaling indices for the gravitational curvature spectrum 
$ \gamma \simeq 2 $ and $ s \simeq 1 $ gives roughly $ n_s \simeq 1.1 $ \cite{hyu18}.  
There are however a number of uncertainties through this derivation, from the evaluation 
of the true values of $ \nu $, $ \gamma $ and $ s $, to normalizing the full spectra.  
It will require more work in the future to narrow down the exact uncertainties 
in these various steps.  
From our first-order calculations, we arrive at a value for 
$ n_s $ within $ \sim 15 \% $ of the measured value from Planck \cite{planck18} 
and in spite of these uncertainties, which seems reasonable for a very first attempt. 

Despite the crude discrepancy at this stage for $ n_s $, which presumably can be 
improved with more precise studies, there are a number of advantages for this 
gravitationally motivated picture, compared to the inflation motivated one.  
The gravitational scenario does not require the postulate of an undiscovered 
quantum field and its fluctuations, but simply utilizes the quantum fluctuations 
of the existing gravitational field, which most current discussions neglect.  
Secondly, unlike inflation models, the theory of quantum gravity is at least in principle unique.  
The number of tunable parameters is extremely minimal and, most of them, 
in principle can be nailed down with further studies.

It should also be pointed out that the far left data point 
(at $ k = 2.9 \times 10^{-4} \, h \, \rm{Mpc}^{-1} $) was released by Planck \cite{planck18} only 
around one week after publication of our previous paper \cite{hyu18}, suggesting a 
reduction in power on the very low $ k $, and thus somewhat supporting the 
quantum gravity prediction.  
It is obvious that the significant magnitude of the error bars for this last point makes 
it vastly premature to make any definite conclusions.  
However, the fact that the data point was published after the publication of the first paper 
could make it a genuine prediction of quantum gravity, instead of a post-diction, 
like the rest of the data points.  
The publication of this new (and previously non-existent) data point suggests the 
predictability and testability of the quantum gravity picture presented here.  
It is hoped that this can be done via new astronomical experiments and observations in the 
near future, which will further improve and narrow down the errors in the small-$ k $, 
and which may serve to verify (or falsify) this theory.

As a final note, it should be noted that the inflation picture requires setting the 
scale of the spectrum with $ P(k) \propto k $ at small $ k $'s.  
It would be difficult for at least basic inflation models to reproduce a dip for $ k \rightarrow 0 $.  
As a result, with future observational experiments, further narrowing of errors bars 
in the low-$ k $ regime may be possible to provide a clear distinction 
between the competing gravitational and inflation picture.  
More detailed comparisons and computations can be found in the previous 
paper \cite{hyu18}.  
In summary, compared to inflation, the gravitationally 
motivated picture provides in our view a competing alternative explanation 
for the power spectrum in terms of both its naturalness and its uniqueness.

\vskip 20pt
	
\section{Constraints on the Scaling Dimension $ \nu $ from Cosmology}
\label{sec:constrainingNu} 
	
\vskip 10pt
	
It should by now be apparent that the universal scaling index $ \nu $ plays an 
important role in the theory of quantum gravity.  
Sec. \ref{sec:introQG} already summarized various methods, both analytical and numerical, 
to determine this value.  
Furthermore, Sec. \ref{sec:QGexpofPk} described a method to relate this to cosmological 
observations, showing that the numerically derived value is in generally good 
agreement with recent observational data.  
It is therefore of interest to see if the logic can be reversed - by taking advantage of the 
variance in the data, to provide a constraint on this important theoretical parameter $ \nu $.
	
\begin{figure}
\begin{center}
\includegraphics[width=0.90\textwidth]{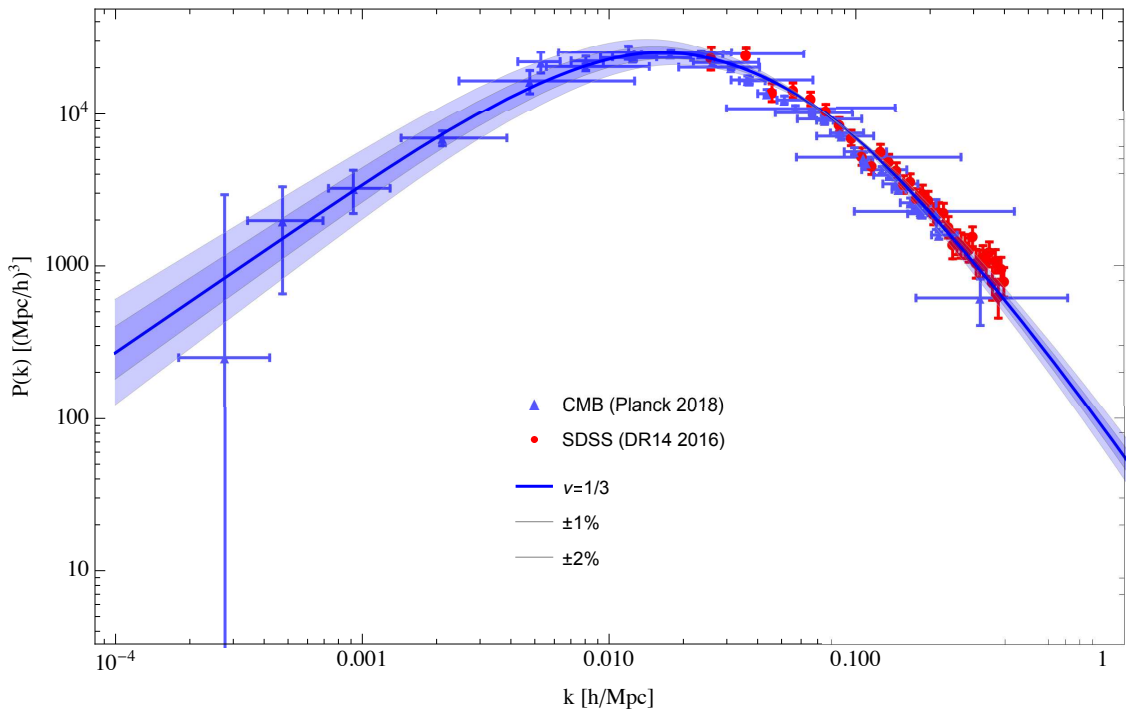}
\end{center}
\caption{Matter power spectrum $ P(k) $ with various choices for the scaling exponent $ \nu $.  
The middle (blue) curve shows the matter power spectrum as predicted by quantum gravity 
with a value of $ \nu=1/3 $ as before, with two bands showing variance of $ 1-2\% $ 
in the value of $ \nu $.  
Notice that in order to obtain general consistency with 
current CMB data, $ \nu $ cannot deviate by more than $ \sim2\% $ from the 
theoretically predicted value of 1/3. This can be viewed as a rather stringent 
constraint on the value of the exponent $ \nu $ arising from cosmology.}
\label{fig:Pubplot3_varynu}
\end{figure}

Fig. \ref{fig:Pubplot3_varynu} shows the same plot as in Fig. 2 (ignoring the effect of IR 
regulation and RG running of Newton's constant $ G $ that is only important in the last 
three data points) but with 
a $1 \%$ and $2 \%$ variance added to the value $ \nu = 1/3 $.  
It can be seen that, should one want to stay within most of the error bars on the left, 
only a maximum of $ 1 \% $ variance in $ \nu $ is allowed.  
A $ 2 \% $ variance would already significantly protrude away from the rather 
stringent vertical error bars in the $ 4^\text{th} $ point from the left.
% Figure 3
One can therefore conclude that current cosmological data provides a very stringent 
constraint on the theoretical value of the scaling exponent $ \nu $ - supporting 
the value of $ \nu \simeq 1/3 $, with a maximum allowed deviation of $ 1-2 \% $.

\vskip 20pt
	
\section{Constraining the Running of $ G $ from Cosmology}
\label{sec:constrainingG} 
	
\vskip 10pt
	
A similar study can be performed for the magnitude of the RG running of $G$, and specifically
the key quantum amplitude $ c_0 $.  
Recall that the running of $G$ is given by 
\begin{equation}
G(k) = G_0 \left[ \;
1 + 2 c_0 \left( \frac{m^2}{k^2+m^2} \right)^{3/2} + 
\mathcal{O} \left( \left( \frac{m^2}{k^2+m^2} \right)^{3} \right)
\; \right]  \;\; ,
\label{eq:Grun}
\end{equation}
where $G_0$ is the currently established laboratory value for Newton's constant, 
the quantum amplitude is
$ 2 \, c_0 \approx 16.04 $, and the nonperturbative gravitational condensate scale is 
estimated at $ \xi \equiv m^{-1} \sim \sqrt{3 / \lambda} \simeq 5300$ Mpc.
The value of $ 2 \, c_0 \approx 16.04 $ is computed from the Regge-Wheeler 
lattice formulation of quantum gravity \cite{ham15}.  
This is largely in exact analogy, both in concept and in practice (via the lattice), 
to the evaluation of the $\beta (g) $ function in QCD.  
The latter represents a quantity that has been extensively tested in collider experiments, 
and is by now in extremely good agreement with accelerator experiments.  
At this stage, unlike $ \beta (g) $ from QCD, the same level of precision has not yet been 
achieved for $ c_0 $, and it seems possible at this stage, given various numerical 
uncertainties inherent in the calculation of $ c_0 $, to have deviations that could 
modify it by up to an order of magnitude.

\begin{figure}
\begin{center}
\includegraphics[width=0.90\textwidth]{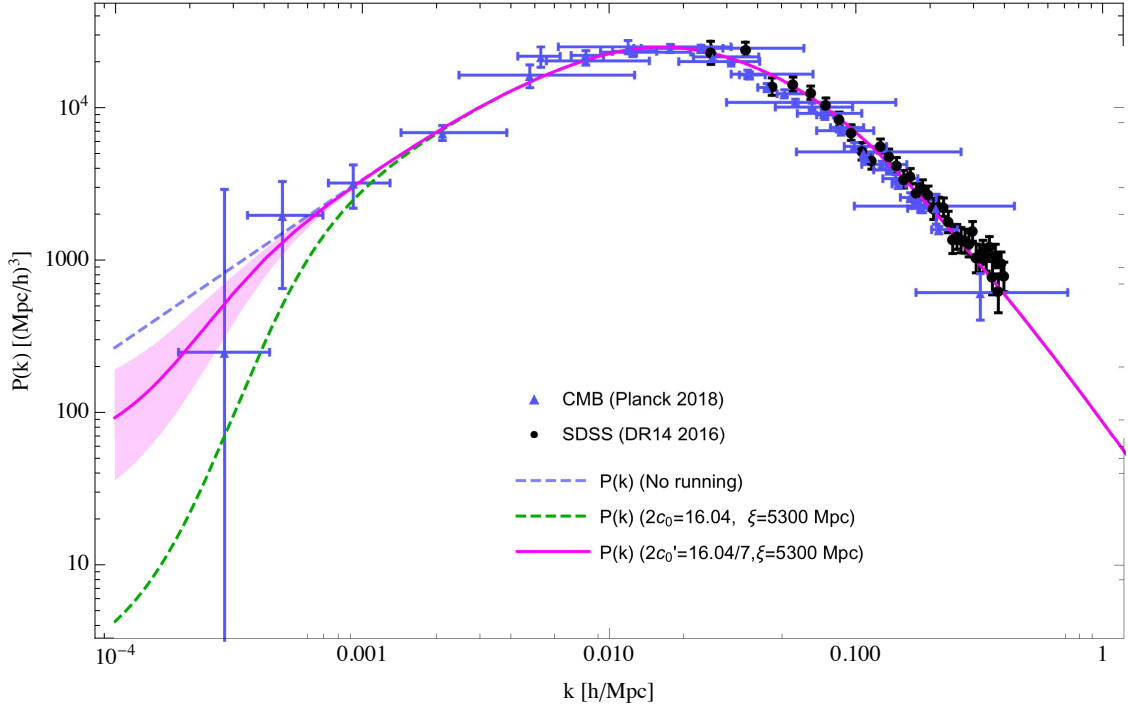}
\end{center}
\caption{Matter power spectrum $ P(k) $ shown for various choices of the quantum 
amplitude $ c_0 $.  
The middle (purple) solid curve shows a best fit through the last few low-k data points via an RG running of Newton's constant $ G $, with an amplitude of running 
$ 2 \, c_0 \approx 16.04/7 \approx 2.29 $, i.e. around $ 15\% $ 
of the preliminary value of 16.04 from the lattice.  
The shaded (purple) band represents a variance in $ 2 \, c_0 \approx 2.29 $ by a factor of 2 and 1/2. 
The original spectrum with no running (top, blue, dashed) and the spectrum 
with running of Newton's constant $ G $, with the original coefficient of 
$ 2 \, c_0 \approx 16.04 $, (bottom, green, dashed) are also shown for reference. 
Note that the middle (purple) curve with $ 2 \, c_0 $ modified to $16.04/7$
can also be mimicked by instead tuning the nonperturbative scale 
$ \xi $ to $ \sim 2.5 \times 5300 \, \rm{Mpc} $ ($ \approx 13,000 \, \rm{Mpc} $) 
and keeping the quantum amplitude $ 2 \, c_0 = 16.04 $.}
\label{fig:Pubplot4_varyc0}
\end{figure}

As a result, one can parallel the previous study of $ \nu $, and utilize cosmological 
data to provide a best-fit value, and thus a constraint, on the quantum amplitude $ c_0 $.  
Also, the same type investigation for the variance in $ \xi $ will be done at the end of this section.
Fig. \ref{fig:Pubplot4_varyc0} shows the best fit to $c_0$, which corresponds to
roughly $ 1/7 $-th, or $ 15 \% $, of the original value for $ c_0 $, 
i.e. $ 2 \, c_0 = 16.04/7 \simeq 2.29 $ (in solid purple).  
The bands above and below the solid purple curve represent a further factor of 
$ 1/2 $ and $ 2 $ respectively.
% Figure 4
Note that the middle solid (purple) curve with $ 2 \, c_0 $ modified to $ 16.04/7 $ can also 
be mimicked by instead tuning $ \xi $ to $ \sim 2.5 \times 5300 $ Mpc ($ \approx 13,000 $ Mpc)
and keeping the lattice value for the amplitude at $ 2 \, c_0 = 16.04 $.   
The initial association of $ \xi \sim \sqrt{3/\lambda} \approx 5300 $ Mpc is theoretically 
motivated by connecting the curvature vacuum condensate scale in the theory, 
$ \sqrt{3/\lambda} $, 
to the nonperturbative correlation length $ \xi $ \cite{loops}, 
and again in close analogy to what happens in QCD
(the factor of $ 1/3 $ is often accompanied with $ \lambda $ in the equations of motions, 
such as the classical Friedman equations).  
It is thus conceivable that the order of estimate $ \xi $ can be varied by a factor 
up to an order of magnitude.  
The above analysis shows that, if the lattice value of $ 2 \, c_0 = 16.04 $ is to be 
taken rigorously, then an increase of $ \sim 2.5 $ on the vacuum scale $ \xi $ would best fit the data.
In essence, the RG running of Newton's constant $ G $ requires two parameters, 
the quantum amplitude $ c_0 $ and the correlation length $ \xi $, to fully determine its form.  
The former is in principle calculable from the lattice, while the latter is best associated 
with the scale $ \sqrt{ 3/\lambda } $ provided by the theory, which determines the long-distance 
decay of Euclidean curvature correlation functions at a fixed geodesic distance.
Nevertheless, the error bars in the last data point in Fig. 4 are too wide to provide 
any definite conclusions at this stage.  
It is conceivable that further satellite experiments might put further constrains on the 
errors in these points, and thus provide more insight on these fundamental microscopic parameters.
	
Here we note that it is of some interest to explore analytical (as opposed to numerical) 
methods related to the running of Newton's constant $ G $.  
One possibility briefly mentioned earlier is the $ 2 + \epsilon $ expansion \cite{wei79, gas78, eps}, 
which provides an estimate for the scaling exponent $ \nu^{-1} $ to be between 2 and 4.4,
through a one- and two- loop double expansion (in $G$ and the dimension) respectively, giving 
additional confidence in the numerically computed value $ \nu^{-1} \simeq 3 $. 
Similar estimates for the exponent $\nu$ are found within a set of truncated RG equations, 
directly in four dimensions 
% \cite{reu98}-\cite{gie15}.
[43-48].
	
\begin{figure}
\begin{center}
\includegraphics[width=0.90\textwidth]{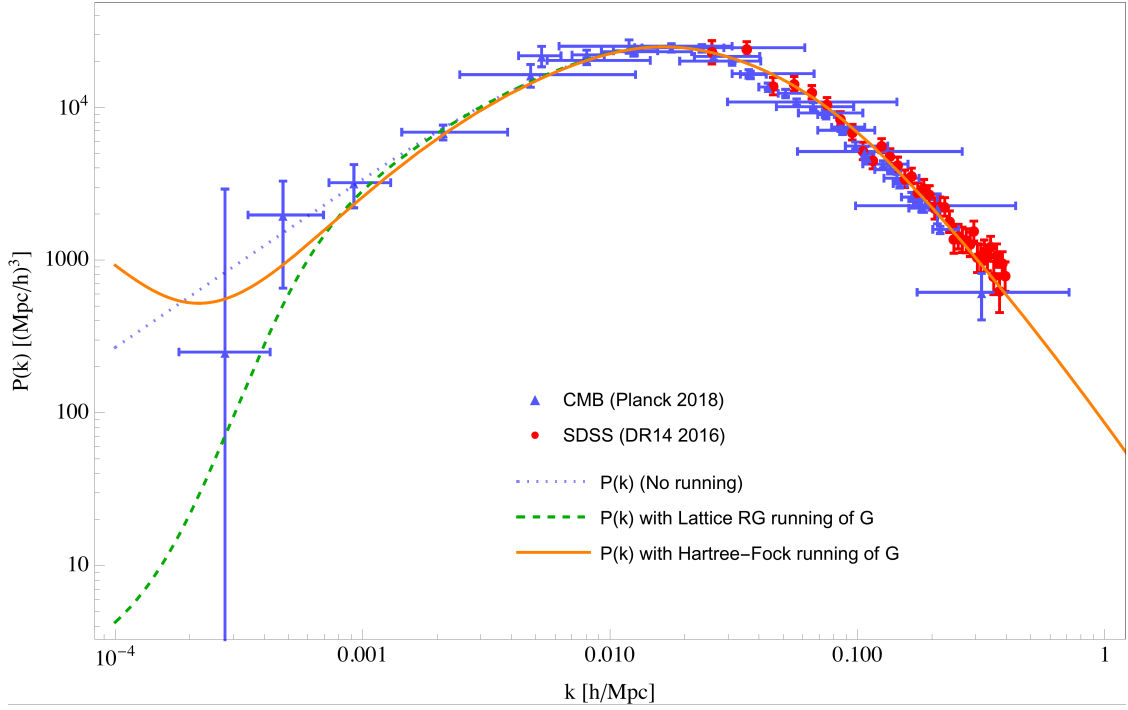}
\end{center}
\caption{Lattice versus Hartree-Fock running of Newton's constant $ G $.  
The middle solid (orange) curve shows $ P(k) $ implemented with the Hartree-Fock 
running of Newton's constant $ G $ factor.  
The lower dashed (green) curve shows the original lattice RG running of 
Newton's constant $ G $ (with the original lattice 
coefficient $ 2 \, c_0 = 16.04 $) for comparison.  
The original spectrum with no running is also displayed by the top 
dotted (blue) curve for reference.}
\label{fig:Pubplot5_pkrunhf}
\end{figure}

Another recently explored idea is a nonperturbative approach via a mean field approximation, 
which in this context is essentially the Hartree-Fock (HF) self-consistent method applied 
to quantum gravity \cite{hyu19hf}, used here for the running of Newton's constant $ G $.  
One finds the following expression for the running of $G$,
\begin{equation}
G \,\, \rightarrow \,\,G_\text{HF}(k) = G_0 \left[
1 -  \frac{3m^2}{2k^2}  \log \left( \frac{3m^2}{2k^2} \right) \right]	
\label{eq:promoteGHF}
\end{equation}
The result of this exercise is shown in Fig. \ref{fig:Pubplot5_pkrunhf}.  
The middle solid orange curve shows the Hartree-Fock expression for the running of 
Newton's constant $ G $, while the bottom dashed green curve and the top blue dotted 
curve show the original lattice running of Newton's constant $ G $ 
(with original lattice coefficient $ 2 \, c_0=16.04 $), as well as no running respectively for reference.  
It seems that the Hartree-Fock running of $G$ is in good consistency with the lattice expression, 
except for the eventual unwieldly upturn beyond $ k < 2 \times 10^{-4} $.  
However, this upturn is most likely an artifact from the Hartree-Fock expression being 
just a first-order analytical approximation after all (it is well known that the 
Hartree-Fock approximation can be
extended to higher order, by including increasingly complex higher loop 
diagrams, with dressed propagators and vertices still determined self-consistently by 
a truncated version of the Schwinger-Dyson equations).
Nevertheless, the Hartree-Fock approximation shows good consistency with both 
the latest available observational  data sets, as well as with the lattice results.  
The fact that it exhibits a gentler dip at small k perhaps also provides support for 
the reduced lattice running coefficient of 
$ 2 \, c_0 = 16.04 / 7 \approx 2.29 $ from Fig. \ref{fig:Pubplot4_varyc0}.
% Figure 5

\vskip 20pt
	
\section{Angular CMB Temperature Power Spectrum}
\label{sec:CmbC_l} 
	
\vskip 10pt
	
The most accurate recent measurements of the CMB are actually 
represented by the angular temperature power spectrum, represented by a set 
of angular Fourier coefficients denoted by $C_l$.  
It is therefore useful to translate the quantum gravity prediction to the angular temperature 
spectrum represented by the $C_l$'s.
The angular temperature spectrum $C_l$ coefficients relate to a two-point correlation 
function of the temperature, when expanded in terms of spherical harmonics labelled
 by $ l $ and $m$.  
The $C_l$ coefficients themselves are defined as
\begin{equation}
C_l  \; \equiv \; \frac{1}{4 \pi}  \int d^2\hat{n}  \int d^2\hat{n}'  \;
L_l (\hat{n} \cdot \hat{n}' )   \, \langle \; \Delta T (\hat{n})  \, \Delta T (\hat{n}') \; \rangle \;\; ,
\label{eq:Cl_def}
\end{equation}
where $ \hat{n} , \hat{n}' $ are two different directions in the sky, and $ L_l (\theta) $ 
the Legendre polynomials.  
Here we avoid the usual common notation, ``$ P_l (\theta) $'', for the Legendre 
polynomials, in order to avoid unnecessary confusion with the various power spectra.
Following \cite{wei08}, fluctuations in the CMB temperature $ \Delta T $ can be expanded 
in plane waves,
\begin{equation}
\left( \frac{\Delta T ( \hat{n} )}{T_0} \right) = 
\int d^3 q \; e^{ i \mathbf{q} \cdot \hat{n} \, r(t_L) }
\left( \, 	F_1(q) + i \hat{q} \cdot \hat{n} \; F_2 (q)\, \right) \;\; ,
\label{eq:DeltaT_planew}
\end{equation}
where $ T_0 = 2.725 $ K (the average CMB temperature measured today),
$ t_L $ the time of recombination, and $ F_{1,2}(q) $ form factors given by
\begin{equation}
F_1 (q) = 
- \frac{1}{2} a^2 (t_L) \ddot{B}_q (t_L) - \frac{1}{2} 
a (t_L) \dot{a} (t_L) \dot{B}_q (t_L) 
+ \frac{1}{2} E_q (t_L) + \frac{\delta T_q (t_L)}{\bar{T} (t_L)}  \;\; ,
\label{eq:F1_general}
\end{equation}
\begin{equation}
F_2 (q) = - q \left(
\frac{1}{2} a (t_L) \dot{B}_q (t_L) + 
\frac{\delta u_{\gamma q} (t_L)}{ a (t_L)} \right) \;\; ,
\label{eq:F2_general}
\end{equation}
The $ B $ and $ E $ functions describe suitable decompositions of the metric perturbations, 
and $ \delta u_\gamma $ is the velocity potential for the CMB photons.  
These form factors simplify for certain gauge choices.  
In the synchronous gauge, $ E=0 $, whereas in the Newtonian gauge 
$ B=0 $ and $ E = 2 \Phi $, which then gives
\begin{equation}
F_1 (q) = \Phi_q (t_L) + \frac{\delta T_q (t_L)}{\bar{T} (t_L)} \;\;  ,
\label{eq:F1_Newt}
\end{equation}
\begin{equation}
F_2 (q) = -  \frac{\delta u_{\gamma q} (t_L)}{ a (t_L)}  \;\; .
\label{eq:F2_Newt}
\end{equation}
The functions $ \Phi $ and $ \delta u_{ \gamma } $, as well as the scale factor $ a(t) $ and 
$ T(t) $, can all be obtained as solutions of the classical Friedmann 
equations combined with
the Boltzmann transport equations, as is done in standard cosmology, 
which will then in principle lead to unambiguous predictions for the  $ C_l $'s.
Note that $ F_1 (q) $ and $ F_2 (q) $ are referred to as ``$ F(q) $'' and ``$ G(q) $'' 
respectively in \cite{wei08}.
Here we will use the former in order to avoid confusion with the expression for the 
running of Newton's constant $ G(k) $, as it will be implemented below. 
We also make the usual approximation of a sharp transition at $ t_L $ during recombination, 
which is quite acceptable since we are primarily interested in the general trend, and not 
exceedingly precise features, of the spectrum at this stage.
	
Perturbations in the above form factors are fully governed by the classical Friedmann and 
Boltzmann transport equations.  
These lead to standard solutions in terms of transfer functions 
$ \mathcal{T}(\kappa) $, $ \mathcal{S}(\kappa) $ and $ \Delta(\kappa) $, given by
\begin{equation}
F_1 (q) = \frac{ \mathcal{R}_q^0 }{ 5 } 
\left[  
3 \, \mathcal{T} \left( 
\frac{q \, d_T }{a_L} \right) R_L \,\, - \,\,  (1 + R_L)^{ -\frac{1}{4} } 
\; e^{ - \left( \frac{q \, d_D }{ a_L } \right)^2 } 
\mathcal{S} \left( \frac{q \, d_T }{a_L} \right)
\; \cos \left[ { \frac{q \, d_H }{ a_L } + \Delta \left( \frac{q \, d_T}{a_L} \right) }  \right]
\right]   \;\; ,
\label{eq:F1_sol}
\end{equation}
\begin{equation}
F_2 (q) = \sqrt{2} \; \frac{ \mathcal{R}_q^0 }{ 5 } \;
%\left[  
(1 + R_L)^{ -\frac{3}{4} } \; e^{ - \left( 
\frac{q \, d_D }{ a_L } \right)^2 } \mathcal{S} \left( \frac{q \, d_T}{a_L} 
\right)	
\; \sin \left[ { \frac{q \, d_H}{a_L} + \Delta \left( \frac{q \, d_T}{a_L} \right) }  \right] 
%\right] \;\; ,
\label{eq:F2_sol}
\end{equation}
where $ a_L =  a(t_L) = 1/(1+z_L) $, $ z_L = 1090 $, $ d_T = 0.1331 $ Mpc, 
$ d_H = 0.1351 $ Mpc, $ d_D = 0.008130 $ Mpc, 
$ d_A = 12.99 $ Mpc, and 
$ R_L \equiv 3 \Omega_B (t_L) / 4 \Omega_\gamma (t_L) = 0.6234 $.  
It is noteworthy at this stage that all three transfer functions are completely determined 
again by (well measured) cosmological parameters.
So the only remaining ingredient to fully determine the $ C_l $ coefficient is an 
initial spectrum $  \mathcal{R}_q^0 $, which is usually parameterized by an 
amplitude $ N $ and spectral index $ n_s $,
\begin{equation}
\mathcal{R}^0_q = 
N \, q^{-3/2} \;  {\left( \frac{q}{ q_\mathcal{R} } \right) }^{(n_s-1)/2} \;\;  .
\label{eq:Rq_param}
\end{equation}
Here the reference ``pivot scale'' is usually taken to be 
$ q_\mathcal{R} = 0.05 \, \rm{Mpc}^{-1} $ by convention.  
As a consequence, once the primary function $ \mathcal{R}_q^0 $ is somehow determined, 
classical cosmology then fully determines the form of the $C_l$ spectral coefficients.
It is therefore possible to write the $ C_l $'s fully, and explicitly, 
in terms of the primary function $ \mathcal{R}_q^0 $.  
After expanding the original plane wave factor in a complete set of spherical Harmonics 
and spherical Bessel functions, one obtains
\begin{equation}
C_l = 16 \pi^2 \, T_0^2 \int_{0}^{\infty}{q^2 \,dq} \;
\left( \mathcal{R}_k^0 \right)^2
\left[ \, 
j_l ( qr_L ) \widetilde{F_1} (q) + j_l^\prime ( qr_L ) \widetilde{F_2} (q) \, 
\right]^2 \;\;  .
\label{eq:Cl_jl}
\end{equation}
Here $r_L = r\left(t_L\right) $,  and we have factored out the function
$ \mathcal{R}_q^0 $ explicitly 
$ F_1 (q) = (\mathcal{R}_q^0) \, \widetilde{F_1} (q) $ 
and 
$ F_2 (q) = (\mathcal{R}_q^0) \, \widetilde{F_2} (q) $.  
Recall that, since the matter power spectrum is given by
\begin{equation}
P \left( k \right) = C_0 \left( \mathcal{R}_k^0 \right)^2 \, k^4 
\left[ \mathcal{T} ( \kappa ) \right]^2 \; ,
\label{eq:Pk_R}
\end{equation}
we can use $ \mathcal{R}_q^0 $ to obtain a direct relation between the 
$C_l$'s and $ P(k) $,
\begin{equation}
C_l  =  16 \pi^2 \; T_0^2 \int_{0}^{\infty} q^2 dq \, P(q) \; 
\left[ \, C_0 \, k^4 \, { \mathcal{T} ( \kappa ) }^2 \, \right]^{-1}
\left[ \,
j_l ( qr_L ) \widetilde{F_1} (q) + j_l^\prime ( qr_L ) \widetilde{F_2} (q) \, 
\right]^2 \,\, ,
\label{eq:Cl_Pk}
\end{equation}
where $ q $ and $ k $ are related by $ q=a_0 k $, and the scale factor ``today''
$ a_0 $ can be taken to be $ 1 $.

\begin{figure}
\begin{center}
\includegraphics[width=1.00\textwidth]{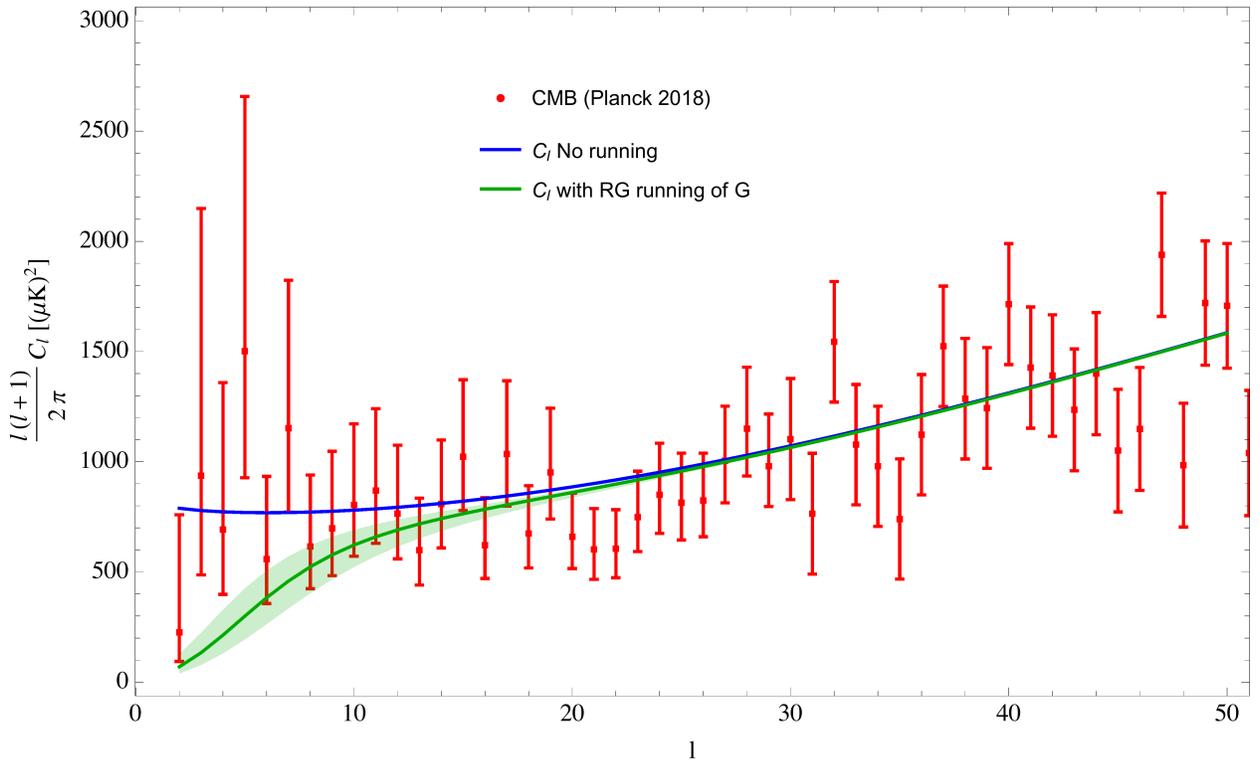}
\end{center}
\caption{Plot of the angular power spectrum coefficients $ l(l+1)C_l/2\pi $.  
The upper (blue) curve shows the quantum gravity prediction of the angular CMB power spectrum 
$ C_l $'s as obtained from the matter power spectrum $ P(k) $ 
(and thus with scaling exponent $\nu$=1/3), without any IR regulation effect from 
$ \xi $, and without the RG running of Newton's constant $ G $.  
The bottom (green) curve shows the combined quantum gravity effect 
now with IR regulation and the lattice RG running of Newton's constant $ G $, 
with the original lattice quantum amplitude $ 2 \, c_0 = 16.04 $.  
The upper and lower bands on the bottom curve represent a factor of 2 variance 
on the quantum amplitude $ c_0 $, i.e. $ 2 \, c_0=(8.02,32.08) $.}
\label{fig:Pubplot6_Cl}
\end{figure}

The quantum theory of gravity, as outlined in the earlier sections, predicts the 
form of the full matter power spectrum function $ P(k) $.  
Using Eq.~(\ref{eq:Cl_Pk}), one can therefore translate the quantum gravity prediction 
on $ P(k) $ to the angular coefficients $C_l$'s.  
Fig. \ref{fig:Pubplot6_Cl} shows a plot of the ensuing result, represented by the top 
blue curve, for $ l = 2 $ to $ l = 50 $.  
One can see that the theoretical prediction (obtained here by numerical integration) 
is in generally rather good agreement with most of the observational data.
% Figure 6
Again, it should be emphasized here, again, that reproducing the full expression for the 
$C_l$'s does not require the inclusion of a scalar field anywhere.  
Instead, the spectrum for gravitational fluctuations is used to set the scaling in a 
particular regime, which is chosen to be the galaxy regime for its most direct connection, 
and the rest is then fully governed by classical general relativity and standard kinetic theory.
Another way of expressing this result is that the entire expression for the $C_l$'s, or for $ P(k) $, 
except for the spectral index $ n_s $ and the amplitude $ N $, 
is fully governed by classical general relativity and kinetic theory 
(and by finely measured cosmological parameters such as 
$ \Omega_M $, $ \Omega_\Lambda $, $  H_0 $, etc. ...).  
That is, $ n_s $ and $ N $ are the only two remaining theoretically undetermined quantities
in this framework.  
Whereas inflation provides one perspective on how these two parameters can be derived, 
quantum gravity provides in our view  an equally well-motivated alternative.

However, as before, additional quantum gravity effects are expected to manifest themselves 
at very large distances comparable to $ \xi $.  
In angular space, this corresponds to the widest angles, or very low-$ l $ regime.  
In this context one can then investigate how the IR regularization and the RG running 
of Newton's constant $ G $ affects the standard prediction, 
thus providing potentially testable predictions and alternatives,
to distinguish between this quantum gravity fluctuation picture and the inflation one.
In the case of the matter power spectrum $ P(k) $, the RG running of 
Newton's constant $ G $ was implemented by modifying
\begin{equation}
P(k) \,\, \rightarrow \,\, \left[ \frac{G_0}{G(k)} \right]^2 P(k)
\label{eq:Pk_runpromote}
\end{equation}
where $ G_0 $ is the Newton's gravitational constant measured in the laboratory or on
solar system scales.  
In the angular spectrum coefficients $C_l$, this will introduce an extra factor of 
$ \left[G_0 /  G(q)\right]^2 $ in the integrand.  
The resulting modification is shown by the lower green curve in Fig. \ref{fig:Pubplot6_Cl}.  
As for the case of $ P(k) $, the RG running of Newton's constant $ G $ causes a 
significant drop in the magnitude of the $C_l$'s at large distance scales (low $ l $).  
The green bands around the curve with the RG running of Newton's constant $G$ shows 
the effects of varying by factor of 2 the quantum running amplitude $ c_0 $.

Note that in particular the last point at $ l=2 $, which corresponds to measuring the 
CMB on the largest scales on the sky, is significantly below the classical prediction, 
and has represented a well-known anomaly for quite some time.  
Although that last point is plagued with large uncertainties due to cosmic variance -
the lack of independent samples on this scale from our sky - many do agree that 
the error bars as shown already account for our best assessments of the associated variances.  
If these judgements are believable, then the classical prediction seems just marginally 
consistent with the allowed uncertainties.
If the matter power spectrum indeed originates from inflation, then there are currently no 
widely agreed solutions that can reasonably explain the sudden drop in power 
at the extreme low $ l $'s. 
This is a general consequence of inflation, providing a scale-invariant normalization 
at very large scales.

\begin{figure}
\begin{center}
\includegraphics[width=1.00\textwidth]{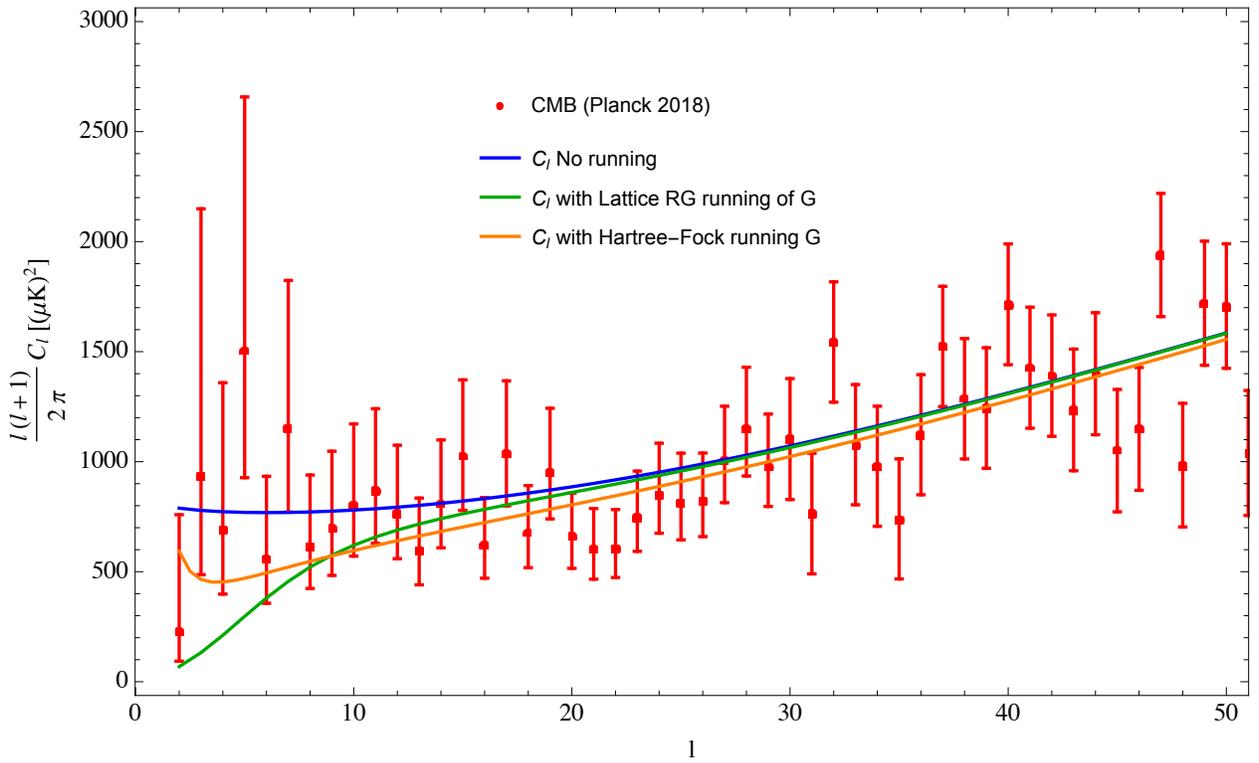}
\end{center}
\caption{
A comparison of the lattice versus Hartree-Fock RG running of Newton's constant $ G $ on the angular spectrum coefficients $ C_l $. 
The middle (orange) curve shows $ C_l $ implemented with a Hartree-Fock running of 
Newton's constant $ G $ factor, in comparison with the original lattice RG running of Newton's 
constant $ G $ (with the original lattice coefficient $ 2 \, c_0 = 16.04 $), represented by the 
lower (green) curve.  
The original angular spectrum with no running of $G$ is shown by the upper (blue) curve 
for reference.
}
\label{fig:Pubplot7_Clrunhf}
\end{figure}

Quantum gravity, however, tells us that the effects from a running Newton's constant $ G $ 
must be included, whether by an expression calculated from the lattice approach as 
represented by the green curve above, or by one calculated via the Hartree-Fock approximation.  
Fig. \ref{fig:Pubplot7_Clrunhf} shows a comparison between them.

So it seems that for distance scales roughly $ r < \xi $, both quantum gravity and inflation produce a 
spectrum that agrees rather well with observations.  
Although one can argue the gravity induced perspective is more natural to the principle of 
Ockham's Razor, being able to explain the same physical phenomena without the need 
of a new field, the question of which picture is more desirable remains 
largely a philosophical one.  
However, for distance scales roughly $ r > \xi $ (i.e. small $ k $ or small $ l $), 
both the observed matter power spectrum $ P(k) $ and the corresponding 
observed angular power spectrum $ C_l $ 
seem to hint towards the quantum gravity picture.  
Of course, ultimately, much more precise data will be needed to conclude this decisively.  
Nevertheless, the current context presents an intriguing possibility for a 
new explanation for the nature of correlations and for the origin of cosmic fluctuations, 
and also beyond that an interesting testing ground for quantum gravity.
% Figure 7

It is of some interest here to compare the running of Newton's constant $ G $ obtained from the lattice to the analytical result of the Hartree-Fock approximation.  
Using the Hartree-Fock expression Eq. (\ref{eq:promoteGHF}), the corresponding result 
is displayed by the orange curve in Fig. \ref{fig:Pubplot7_Clrunhf}.   
One notes that, similarly to $ P(k) $, the Hartree-Fock expression analogously 
(i) predicts a smaller power in the low-$ l $'s ($ l < 50 $), (ii) has a less dramatic dip
compared to the lattice running at the very large scales ($ l < 10 $), and 
finally (iii) predicts a somewhat unwieldly upward turn at the extreme large scales ($ l < 3 $).  
As is the case of the matter power spectrum $ P(k) $, some of these features, 
especially the unwieldy upturn at extremely large scales, may be an artifact from the fact 
that Hartree-Fock is essentially a mean-field type approximation.  
Nevertheless, while the lattice prediction may be more trust-worthy due to it being an 
exact, numerical evaluation of the path integral, the Hartree-Fock expression provides a 
good independent consistency check for this picture.

\begin{figure}
\begin{center}
\includegraphics[width=1.00\textwidth]{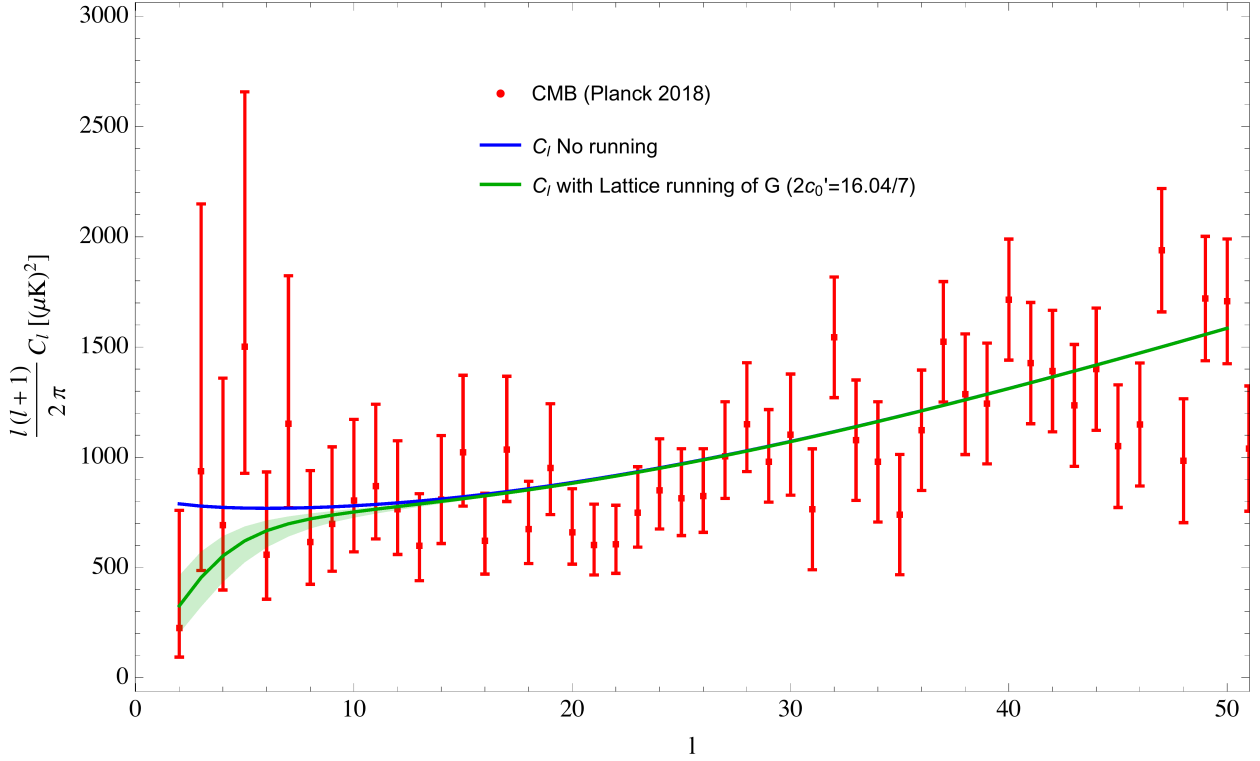}
\end{center}
\caption{
Angular power spectrum shown, with a comparison between various choices for the lattice 
RG running of $ G $ parameters of Eq.~(\ref{eq:Grun2}).  
For reference, the top blue curve represents the original spectrum with no RG running 
of Newton's constant.  
The bottom green curve shows the effect of the lattice RG running of Newton's constant 
$ G $ with a modified value for the lattice amplitude $ 2 \, c_0 = 16.04 / 7 \approx 2.29 $.  
This new curve, represented by the modified amplitude $ 2 \, c_0$, appears to fit best 
through the last few ($l < 10$) data points.  
Here the green bands represent a factor of two in variance around this modified $ c_0 $.
The last curve reveals that although the original value of $ 2 \, c_0$, as obtained from 
numerical lattice simulations,  is around the correct order of magnitude,
nevertheless when looked at more carefully, a slightly smaller value seems to be 
favored by the very low $l$ cosmological data.  
Note that a coefficient of $ 2 \, c_0 \approx 16.04 / 3.13 $ will allow the green curve to 
precisely go  through the last point at $ l=2 $.  
However cosmic variance suggests $\Delta C_l \sim 2 /\sqrt{2l+1} $, which disfavors
giving too much weighting to the final point.
}
\label{fig:Pubplot8_Cl}
\end{figure}

Finally, it is possible to investigate the effect of varying the lattice quantum amplitude $ c_0 $
appearing in the running of $G$, as in Eqs.~(\ref{eq:Grun1}) and (\ref{eq:Grun2}).  
From the investigation of $ P(k) $, the value of $ c_0 $ that best fits the large scale data 
at small $k$ is $ 2 \, c_0 = 16.04 / 7 \approx 2.29 $.  
Fig. \ref{fig:Pubplot8_Cl} plots the effect of this modification.  
As before, this choice seems to fit rather well with most of the data in the low-$ l $ regime.  
Nevertheless it should be noted that this modification can also be mimicked by modifying 
the correlation length $ \xi \approx 5300 $ Mpc to $ \xi \approx 13000 $ Mpc, 
or by any combined adjustments of the two parameters $\xi$ and $c_0$.
% Figure 8
Although at first sight it may seem impossible to eventually distinguish the difference 
between the classical and quantum picture by the still highly uncertain data, 
it may not be so with better telescopes in the near future.  
Table \ref{tab:Clcompare} shows the percentage difference between the classical prediction, 
and one with the RG running of Newton's constant $G$ included, in accordance with 
Eq.~(\ref{eq:Grun2}), with the choice $ 2 \, c_0 = 16.04 / 7 \approx 2.29 $.  
\begin{table} 
\begin{center} 
\begin{tabular}{|l|l|l|l|l|} 
\hline\hline
$ l $ & $ l (l+1) C_l / 2\pi   $ Classical & $ l (l+1) C_l / 2\pi   
$  Quantum  & Difference  &  $ \% $-Difference
\\ \hline \hline
2 &   788.8  & 328.5   & 460.3   & 58.4 $ \% $
\\ \hline 
3  &  778.4  &  457.4  &  321.0  &  41.2  $ \% $
\\ \hline 
4  &  772.6  &  555.4  &  217.2  &  28.1  $ \% $
\\ \hline 
5  &  769.5  &  623.2  &  146.3  &  19.0  $ \% $
\\ \hline 
6  &  768.6  &  668.6  &  100.0  &  13.0  $ \% $
\\ \hline 
7  &  769.3  &  699.5  &  69.8   &  9.07  $ \% $
\\ \hline 
8  &  771.6  &  721.6  &  50.0   &  6.48  $ \% $
\\ \hline 
9  &  775.2  &  738.5  &  36.7   &  4.73  $ \% $
\\ \hline 
10  &  780.0  &  752.5  &  27.5  &  3.53  $ \% $
\\ \hline \hline 
\end{tabular} 
\end{center} 
%\center{\small {\it 
%TABLE I.  test. 
%\medskip}} 
\caption{Values of, and percentage differences between, the classical predictions for the 
angular spectrum coefficients $C_l$'s, and the prediction with a quantum RG running for 
Newton's constant $G$ included.
The quantum gravity values for the $ C_l $'s were computed here with a lattice RG 
running quantum amplitude of $ 2 \, c_0=16.04/7=2.29 $.} 
\label{tab:Clcompare} 
\end{table}
From Table \ref{tab:Clcompare}, one can see a $ 58 \% $ difference in the $C_l$ for $ l=2 $, 
which future experiments may be able to distinguish.  
However, too much emphasis should not be put in the extreme low-$ l $ points 
due to statistical limitations arising from cosmic variance, 
which under reasonable assumptions (mainly Gaussianity)
is expected to grow rapidly as $l$ decreases,  by $\Delta C_l  \sim 2 /\sqrt{2l+1} $ \cite{wei08}.  
Nevertheless, focusing on the $ l=6 $ to $ l=10 $ points, the narrowing of errors
needed to distinguish between the classical and the quantum predictions may very 
much be achievable.  
Thus, for example, for $ l=6 $, the percentage difference between the predictions is $ 13\% $.  
In comparison, the current errors on the Planck data for the 
$ l=6 $ value is $ {+67\% \atop {-36\%}} $.  
For $ l=10 $, the current uncertainties in the Planck data is $ {+46\% \atop{-29\%}} $, 
whereas the difference between the classical and quantum prediction is only $ 3.53\% $, 
which may be more difficult to resolve with future satellite experiments.  
Nonetheless, the magnitude of the errors in the CMB observational data may 
initially seem unpromising to make any claims, this table shows there may still be 
hope in distinguishing the various theoretical predictions.  
Within the $ l=6 $ to $ l=10$ points range, the improvements in technology needed 
to improve the measurements and support the validity of the gravitational 
fluctuation  picture may actually be within reach in the next decades, 
which provides an exciting prospect for the future.

In conclusion, in this section we showed how the quantum gravitation 
prediction for $ P(k) $ unambiguously translates to a prediction for the $ C_l $'s
- which is essentially related to the former via a spherical Bessel transform, 
weighted by some suitable combination of transfer functions.  
The transfer functions in turn are ultimately just solutions to the classical
Friedmann equations  and associated 
Boltzmann transport equations which, apart from the measured values of 
standard cosmological 
parameters such as $ H_0 $, $ \Omega_m $, etc., require no further theoretical input.  
As a result, we were able to show how the quantum gravity prediction for the matter
power spectrum $ P(k) $ directly and unambiguously translates to the angular
coefficients $ C_l $.  
It can be seen that the prediction is rather consistent with current cosmological data.

We also discussed several theoretical parameters, which in this picture potentially have 
some variance and related uncertainties.  
The first two key parameters in the quantum gravity motivated picture are 
the universal scaling exponent $ \nu $, and the fundamental vacuum condensate 
correlation length $ \xi $.  
A third additional parameter here is the quantum amplitude $ c_0 $, 
which governs the amplitude of quantum correction in the RG running of
Newton's constant $G$ as given in Eqs.~(\ref{eq:Grun1}) and (\ref{eq:Grun2}).
Of the three parameters, as shown above in Sec. \ref{sec:constrainingNu}, $ \nu $ 
is pretty much highly constrained (both theoretically and observationally) around $ \nu^{-1} \simeq 3 $.  
The value of this last parameter should also be the most trustworthy of the three, 
since, as a universal scaling exponent, it is expected to be entirely independent of 
schemes and regularization.  
On the other hand, the values of $ \xi $ and $ c_0 $ are somewhat less definite.
Here the nonperturbative length scale $\xi$ is quite analogous (via the observed 
scaled cosmological constant $\lambda = 3 /\xi^2 $), 
to the vacuum condensate scale in QCD 
$ \langle F_{\mu\nu}^2 \rangle \sim 1 / \xi^4 $, or to the scaling violation parameter 
$ \Lambda_{\bar{MS}} \propto \xi^{-1} $.
Which implies that it's absolute value in physical units is not determined 
theoretically, and can ultimately only be fixed by experiment.
Current cosmological data seem to suggest the best - and most consistent - estimate for 
$\xi$ is roughly $ \xi \simeq 2.5 \times 5300 $ Mpc, whereas for the quantum 
amplitude $ 2 \, c_0 \simeq 16.04 / 7 $, or some degenerate combination between the two
(as discussed earlier).  
Finally, we explicitly listed the percentage differences between the classical prediction and 
the quantum one implemented with an RG running of Newton's $G$, which should provide
a useful guide as to how improved observational data must trend in order to support the
picture advocated here.  
Although such precision has not been achieved yet, it should hopefully be attainable in the 
near future, and thus provide an additional significant test for the quantum gravity picture.

Nevertheless, with these flexibilities in mind, the quantum gravity picture outlined here
provides a radically different perspective to the origin of matter and radiation fluctuations, 
compared, for example, to inflation.  
In addition, the inflation picture normalizes the spectrum at large scales (i.e. small-$ k $), 
so that for the $ C_l $'s it predicts a flat scale-invariant plateau for small-$ l $'s.  
If the last point of $l=2$ is to be taken seriously, despite the flexibility inherent in 
various inflation models,  it would be difficult to account naturally for the reduction in 
power on the very largest scales.  
In contrast, the demand of a renormalization group running Newton's constant in the 
quantum gravity picture appears to explain the dip quite naturally.  
Of course, due to the large uncertainties in the data at small $l$'s, significant improvements 
on the errors needs to be made before definite conclusions can be drawn.

\vskip 20pt

\section{Conclusion}
\label{sec:conclusion}  

\vskip 10pt

In this paper, we have revisited the derivation of the galaxy and cosmological matter power 
spectrum that is purely gravitational in origin, which is to our knowledge is the first of its kind 
without invoking the mechanism of inflation.
We provided updated observational data, including revised experimental errors, and 
outlined an elementary study of the uncertainties involved for the theoretical parameters
in this picture, including the universal scaling exponent $ \nu $, the quantum amplitude $ c_0 $ 
and the nonperturbative scale $ \xi $.  
We also presented the Hartree-Fock results for the running of Newton's constant $ G $, 
in the form of an analytical approximation, for a useful comparison with the Regge-Wheeler
lattice result.  
We then extended our predictions to the angular temperature power spectrum and 
repeated and extended the uncertainty analysis.
In both cases, we showed good general consistency of this purely gravitationally motivated 
picture with current observational data, and pointed out a significant deviation 
from the inflation motivated  predictions in the large distance scale regime.
Although experimental errors and cosmic variance are large in this regime, these results provide
a potentially exciting area that can verify, or falsify, the various pictures. 

To reiterate, the primary benefit of the quantum gravity explanation over inflation is the 
non-necessity for additional and untested physics ingredients, other than standard Einstein's 
gravity and accepted modern quantum field theory methods.  
The basis of the methods begins with the path integral formulation of gravity \cite{fey63, fey95} 
which, unlike inflation, provides a very constrained theoretical framework.  
Also, given the well-known fact that the theory is not perturbatively renormalizable, 
standard nonperturbative methods and approaches must be used.  
While a lot of the results used here are derived from the lattice numerical treatment, 
additional confirmations via analytical methods are also briefly discussed, 
including the 2+$ \epsilon $ and the Hartree-Fock approximation.  
The general consistencies of these numerical and analytical methods gives confidence to the results.  
The lattice treatment in particular has a long history of high precision success in other fields, 
from QCD to condensed matter and statistical systems, and thus provides us with particularly 
trustworthy results. 

On the other hand, for inflation, a new, minimum of one, inflaton field, usually scalar in nature, 
must be invoked.  
The details of the particular theory are also highly flexible, leading to a myriad of models, 
see for example \cite{gut14} and references therein.
In addition, recent studies have shown that a majority of single-field inflation models have either 
been ruled out or highly constrained.  
The amount of flexibility for inflation has thus led many to question the predictability and 
ultimate naturalness of such scalar-field based solutions \cite{ste12, teg05, ste13, ste14, ste17}.  
Although the original model of inflation was invented to explain the flatness and horizon 
problem, it has been convincingly argued that it is not a necessary ingredient to do so  
% \cite{ven02, ven04, ven07, har83, har08, hol02,muk81}.  
[62-68].
We have argued therefore that the gravitational picture provides a more concrete 
and natural explanation to the origin and distribution of cosmological matter fluctuations.  

Finally, we have pointed out that the gravitational fluctuation picture also provides a 
clear set of predictions that diverge from scalar field induced predictions on very large scales.  
As advanced satellite experiments are continuously being conducted, and increasingly 
accurate measurements are becoming available, the predictions originating in quantum 
gravity outlined in the previous section could be verified or disproved in the near future. 

We should add that it is certainly possible for our picture of gravitational fluctuations to 
even coexists with inflation, with both effects providing significant contributions to power spectra.  
Nevertheless, we do not explore this idea in depth here, as the primary aim of this paper is 
to show that the same power spectra can be reproduced purely from macroscopic 
quantum fluctuations of gravity, independent of any inflation mechanism, making use
of well-accepted and tested methods for dealing with perturbatively nonrenormalizable 
theories.  
Still, this could be a potential avenue for future explorations. 

In addition to the results presented here, there are also a number of exciting future 
directions which seem meaningful to explore.  
For example, the quantum gravity-based explanation is most certainly not Gaussian, 
due to the presence of non-trivial anomalous scaling dimensions \cite{ham17} 
which affect all gravitational  $n$-point functions, although they may seem to 
be Gaussian in certain regimes.  
While the corresponding predictions for the two-point functions, or power spectra, 
are similar to those motivated by inflation, a divergence will definitely be expected 
on higher order $n$-point functions, commonly known as bispectra and trispectra in 
the cosmology context.
For example, the two-point function scalar curvature result of Eq.~(\ref{eq:GRscaling}), 
derived from quantum gravity, will also determine the form of the connected 
reduced three-point function, or bispectrum, for large scale 
scalar curvature correlations \cite{ham17}
\begin{equation}
\langle \; R(x_1) \; R(x_2) \; R(x_3) \; \rangle_{c , R}
\;\; \mathrel{\mathop\sim_{d_{ij} \; \ll \; \xi }} \;\;
{ C_{123}  \over d_{12} \, d_{23} \, d_{31}  } \; ,
\label{eq:corr_triple}
\end{equation}
with constant amplitude $ C_{123} $, and relative geodesic distances 
$ d_{ij}=\left|x_i-x_j\right| $ in coordinate space.  
It is easy to see that this last correlation leads to a Fourier transform in 
momentum- or wavenumber-space of the form
\begin{equation}
B_R ( {\bf k}_1, {\bf k}_2, {\bf k}_3 ) \; \equiv 
\;
\langle \, R({\bf k}_1) \, R({\bf k}_2) \, R({\bf k}_3) \, \rangle_{c,R}
\; \mathrel{\mathop\sim_{ k_{i} \; \gg \; m }} \;
\frac { \log \left(  k_1 + k_2 + k_3 \right) + \gamma_E } 
{ k_1 \, k_2 \, k_3}   \;
\delta^{(3)} \left (  {\bf k}_1 + {\bf k}_2 + {\bf k}_3 \right ) \; ,
\label{eq:corr3ft}
\end{equation}
where ${\bf k}_1,{\bf k}_2,{\bf k}_3 $ are the three momenta conjugate to 
$ d_{12}, d_{23}, d_{31} $, the scale $ m = 1 / \xi $, and $ \gamma_E $ is Euler's constant. 
The overall multiplicative constant in Eq.~(\ref{eq:corr3ft}) is 
$\tilde{C}_{123} = C_{123} \times (2 \pi )^3 \times 2 \pi^3 / \Gamma (-9/4) $, with
the expectation that $C_{123} = {\cal O} (1) $ if the curvature two point function 
of Eq.~(\ref{eq:GRscaling}) is normalized to unity (see further discussion below).
One can then follow the same line of argument given in Sec. \ref{sec:QGexpofPk} to 
relate this to measured quantities from the CMB.  
Here we outline the main points of the argument.
Firstly, the Einstein's field equations of Eq.~(\ref{eq:EFE} allow this bispectrum 
for curvature to directly translate to the bispectrum for matter 
$ \langle \, \delta\rho \, \delta\rho \, \delta\rho \, \rangle $.  
Secondly, the transfer function, responsible for turning the two-point spectrum 
from $\sim 1/k$ to $\sim k$ when connecting the late-time galaxy regime to 
the early-time CMB regime, essentially supplies an extra factor of $k$ 
for each fluctuating field.
This then leads to the result
\begin{equation}
B_{\delta\rho}^{\text{(CMB)}} ( {\bf k}_1, {\bf k}_2, {\bf k}_3 ) 
\; \equiv \;
\langle \, \delta\rho ({\bf k}_1) \, \delta\rho ({\bf k}_2) \, \delta\rho ({\bf k}_3) \, \rangle 
\; \mathrel{\mathop\sim_{ k_{i} \; \gg \; m }} \;
\left [ \, \log \left( k_1 + k_2 + k_3 \right) + \gamma_E  \, \right ] \; 
\delta^{(3)} \left (  {\bf k}_1 + {\bf k}_2 + {\bf k}_3 \right ) \; .
\label{eq:corr3rho}
\end{equation}
Nevertheless, most CMB bispectrum measurements are presented nowadays in 
terms of the Bardeen field $\Phi$, which roughly relates (as it describes a specific 
metric component) to the curvature by $ R \simeq \Box \, \Phi$ in the weak field limit.  
This supplies an additional factor of $-k^2$ for each field, giving the following 
explicit prediction for the bispectrum of the $\Phi$ field
\begin{equation}
B_{\Phi}^{\text{(CMB)}} ({\bf k}_1, {\bf k}_2, {\bf k}_3) 
\; \equiv \;
\langle \, \Phi ({\bf k}_1) \, \Phi ({\bf k}_2) \, \Phi ({\bf k}_3) \, \rangle 
\; \mathrel{\mathop\sim_{ k_{i} \; \gg \; m }} \;
f_{NL} \, \cdot \,
\frac{ \log \left( k_1 + k_2 + k_3 \right) + \gamma_E }
{ k_1^2 \, k_2^2 \, k_3^2} \;
\delta^{(3)} \left (  {\bf k}_1 + {\bf k}_2 + {\bf k}_3 \right ) \; .
\label{eq:corr3phi}
\end{equation}
Here the quantity $f_{NL}$ here represents an overall dimensionless amplitude 
for the expected non-Gaussian effects.

However, these non-Gaussian amplitudes are expected to be rather small, 
with suppressions by factors of $1 / \xi$ \cite{ham17}.  
This follows simply from the fact that in real space one has for the semiclassical 
curvature two-point function 
$ \langle \, R \, R \, \rangle \sim 1 / \xi^2 r^2 $, whereas for the scalar curvature
three-point function $ \langle \, R \, R \, R \, \rangle \sim 1 / \xi^3 r^3 $, and
also $ \langle \, R \, R \, R \, R \, \rangle \sim 1 / \xi^4 r^4 $ etc.,  
where $r$ here represents the relevant and appropriate combination of relative 
distances for each reduced curvature $n$-point function.
As a consequence, in Eq.~(\ref{eq:corr3phi}) $ f_{NL} = C_T / \xi^3 $ where 
the remaining amplitude $C_T$ inherits additional transfer function 
parameters from $\kappa$ and $k_{eq}$ as in Eq.~(\ref{eq:Tk}), so
that here $ f_{NL} \sim  1 / ( k_{eq}^3 \, \xi^3 ) \sim 10^{-5} $.
More detailed analyses on this issue, and on the magnitude of these bispectra,
shall be left for future work.  
Nevertheless, it should be clear at this point that analogous results, as hinted above,
can also be derived for various four-point functions.
We note here that the relevance and measurements of such
nontrivial (non-Gaussian) three- and four-point 
matter density correlation functions in observational cosmology were already discussed 
in detail some time ago by Peebles in \cite{pee98}.  
The results presented here imply that further observational constraints on these 
higher order $ n $-point functions could potentially provide additional tests on the 
vacuum condensate picture for quantum gravity as outlined in \cite{ham17}, 
and more specifically the implications of a non-trivial gravitational scaling 
dimensions scenario as described previously.  

In addition, it is clear that the gravitational fluctuation-based explanation presented 
here should also give rise to nontrivial tensor perturbations,
of magnitude comparable to the scalar one.
This could lead to new insights on the corresponding tensor-to-scalar 
ratio parameter $ r $ \cite{ste06}, and to a number of potentially interesting and 
testable consequences to be explored. 
Here we note that tensor perturbation require at first the knowledge of the 
semiclassical Ricci tensor (as opposed to the scalar curvature) correlation functions,
\begin{equation}
\langle \; R_{\mu\nu} (x_1) \; R_{\rho\sigma} (x_2) \; \rangle
\;\; \mathrel{\mathop\sim_{d_{12} \; \ll \; \xi }} \;\;
{ P_{\mu\nu, \rho\sigma}  \over \left ( d_{12} \right )^\Delta} \;\; ,
\label{eq:corr_ricci}
\end{equation}
with polarization tensor $P$ and relative geodesic distance 
$ d_{12}=\left| x_1 - x_2 \right| $ in coordinate space.  
These correlations have not been measured yet on the lattice, but should
be calculable in the near future.
Nevertheless, based on the known scaling dimension for the scalar curvature, 
one would expect here the same result for the operator $R_{\mu\nu} (x) $, 
namely $\Delta =2 $, as in Eqs.~(\ref{eq:GRcorr_scaling}) and 
(\ref{eq:corr_triple}) for the scalar curvature $R$ case.
In turn, these curvature correlations functions can then be related to suitable 
matter and radiation sources, via the quantum equations of motion
\begin{equation}
R_{\mu\nu} (x) \; = \; 8 \pi G \, \left [  \,
T_{\mu\nu} (x) - \half \, g_{\mu\nu} (x) \, T^{\lambda}_{\;\;\lambda} (x) \right ]
\label{eq:tensor_equation}
\end{equation}
with the (trace reversed) $T_{\mu\nu}$ here representing either matter or 
radiation contributions, and thus in complete analogy to what was used earlier 
in Eq.~(\ref{eq:EFE}), and following, for the scalar (trace) case.
Since the scalar curvature correlation function of Eq.~(\ref{eq:GRcorr_scaling})
involves traces of the Ricci tensor (here we make use of the weak field limit) 
$ \langle \left ( R_{00} (x_1)  \!\! + \!\! R_{11} (x_1) \!\! + \!\! R_{22} (x_1) \!\! + \!\! R_{33} (x_1)  \right ) 
\left ( R_{00} (x_2)  \!\! + \!\! R_{11} (x_2) \!\! + \!\! R_{22} (x_2) \!\! + \!\! R_{33} (x_2)  \right ) \rangle $ 
versus say the tensor correlation 
$ \langle R_{12} (x_1) \, R_{12} (x_2) \rangle $, one would expect, based just on
Lorentz symmetry, for the ratio of tensor over scalar correlation amplitudes 
$1/ 4^2 = 1/16 $.
The translation of these simple results into measurable cosmological predictions is
of course a lot more complicated.

In conclusion, the ability to reproduce the cosmological matter power spectrum 
has long been considered one of the``major successes'' for inflation-inspired models. 
Although within our preliminary study, further limited by the accuracy of present 
observational data, it is not yet possible to clearly prove or disprove either idea, 
the possibility of an alternative explanation without invoking the machinery of 
inflation suggests that the power spectrum may not be a direct consequence nor
a solid confirmation of inflation, as some literature may suggest. 
By exploring in more detail the relationship between gravity and cosmological 
matter and radiation, together with the influx of new and increasingly accurate
observational data, one can hope that this hypothesis can be subjected to 
further stringent physical tests in the near future.

\vskip 30pt

\section*{Acknowledgements}
\label{sec:ackn}  

%\vskip 10pt

The authors gratefully acknowledge useful discussions with Stanley Brodsky, 
Jan Hamann, James Peebles, Michael Peskin, Terry Tomboulis, Robert V. Wagoner, 
Risa Wechsler and Yvonne Wong.

% \vskip 30pt

\newpage

\appendix

\section*{Appendix}

\vskip 20pt

\newsection{Magnitude of Quantum Gravity Effects on Solar-System Scales}
\label{sec:solar}
\hspace*{\parindent}

It is of some interest to investigate the magnitude of quantum gravitational effects on 
Solar System scales, and see if they could become potentially significant. 
This paper utilizes three particular results from a quantum treatment of gravity - 
the two-point correlation functions, the infrared (IR) regulator, and the renormalization 
group (RG) running of Newton's constant $ G $.
First, it is easiest to see the IR regulator and RG running of Newton's constant $ G $ play 
completely negligible roles in the Solar System.  
Using 
$ \xi \sim \sqrt{ 3 / \lambda } \simeq 5300 \, \rm{Mpc} = 1.093 \times 10^{15} $ AU, 
the respective modifications
\begin{equation}
\frac{1}{k^2} \rightarrow \frac{1}{k^2+m^2}
\;\;  ,
\label{eq:IRreg3}
\end{equation}
\begin{equation}
G \rightarrow  G + \delta G (k) + \mathcal{O} \left( \delta G^2 \right) 
\,\,\,\,  \text{, where}  \,\,
\frac{\delta G}{G} \equiv 2 \, c_0 \left( \frac{m^2}{k^2+m^2} \right)^{3/2} 
\label{eq:promoteGrun2}
\end{equation}
are only significant when below $ k \simeq m = 1/\xi $, or above $ r \simeq \xi $.  
Taking the Solar System size as $ r_\text{sol} \approx 100 $ AU, 
such quantum effects in the Solar System are suppressed by large 
factors of $ r_{\text{sol}} / \xi \simeq {10}^{-13} $.  
For example, for the running of Newton's constant $ G $, one can estimate 
\begin{equation}
\frac{\delta G}{G} \sim \left( \frac{r_\text{sol}}{\xi} \right)^3 \sim {10}^{-39} \;\; .
\label{eq:running_sol}
\end{equation} 
Next, for the scaling of correlation functions, the fluctuations are governed by the 
Einstein field equations
\begin{equation}
\left\langle \, \delta R \, \delta R \, \right\rangle \; = \; 
\left( 8 \pi G \right)^2  {\bar{\rho}}^2 \, 
\left\langle \, \frac{\delta\rho}{\bar{\rho}}  \,  \frac{\delta\rho}{\bar{\rho}} \, \right\rangle 
\simeq \; G^2 \,  {\bar{\rho}}^2  \, \left( \frac{r_0}{r} \right)^2 \,\, ,
\label{eq:GR_scale2}
\end{equation}
where $ r_0 \simeq 10 \, \rm{Mpc} \sim 10^{-2} \, \xi $, and $ \bar{\rho} $ 
the average matter density of the Universe, which is roughly
\begin{equation}
\bar{\rho} \simeq \frac{M}{\xi^3} \,\, ,
\label{eq:rhobar}
\end{equation}
where $ M $ is of the order of the mass of the currently observable Universe, 
roughly $ M \simeq{10}^{80} $ protons, 
and $ \xi $ is roughly the size the currently observable Universe.  
In the following, again, we are just interested in rough order of magnitude estimates.
The value of Newton's constant $ G $, as argued for ex. in \cite{fey95}, is roughly 
\begin{equation}
G\simeq \frac{\xi}{2 \, M}  \,\, .
\label{eq:Gest}
\end{equation}
So, putting together the numbers one has
\begin{equation}
\left\langle \delta R \, \delta R \right\rangle 
\simeq 
\left( \frac{\xi}{M} \right)^2 \left( \frac{M}{\xi^3} \right)^2 
\left( \frac{ 10^{-2} \, \xi}{r} \right)^2 
\; = \; \frac{1}{\xi^2 \, r^2} \cdot 10^{-4}
\label{eq:RRcorr}
\end{equation}
In a semiclassical approach, one can relate fluctuations in the curvature to metric 
fluctuations  via the weak field relationship
\begin{equation}
\delta R \, \simeq \half  \, \Box \, h \,\, ,
\label{eq:semiclassappr}
\end{equation}
and inserting the value for $ \xi $ then gives
\begin{equation}
\left \langle \; \delta R \, \delta R \; \right \rangle 
\simeq \, \left \langle \; \half \Box h \; \half \Box h \; \right \rangle \,
\simeq \, \frac{10^{-4}}{\xi^2 \, r^2 } \,
\sim \, \frac{10^{-34}}{ (1 \, \text{AU})^2 } \cdot \frac{1}{r^2} \;\; .
\label{eq:GR_sol}
\end{equation}
Therefore, if we use Poisson's equation $ \Box h \simeq \Delta \Phi \simeq 4 \pi G \, \delta \rho $ 
to relate the metric to the matter density in the Solar System, 
it should still obey a $ 1/r^2 $ 
scaling law, but with an amplitude suppressed by a very large factor $ {10}^{-34} $. 
In conclusion, within Solar System scales, any other Newtonian dynamics will completely 
dominate over the (very tiny) correlations due to quantum fluctuations of the 
gravitational field.

\vfill


\newpage

\begin{thebibliography}{99}

% Peebles book

\bibitem{pee93}
P.~J.~E.~Peebles,
{\it Principles of Physical Cosmology},
Princeton Series in Physics (Princeton University Press, NJ, USA, 1993).

\bibitem{pee98}
P.~J.~E.~Peebles,
{\it Issues for the Next Generation of Galaxy Surveys},
{\sl Phil. Trans. Roy. Soc. Lond.} {\bf A357}, 21-34 (1999).

% Galaxy correlations & matt. dens. power spectrum

\bibitem{bah03}
N.~A.~Bahcall et. al.,
{\it The Richness-dependent Cluster Correlation Function: Early Sloan Digital Sky Survey Data},
Astrophys.\ J.\ {\bf 599}, 814 (2003).

\bibitem{bau06}
C.~Baugh,
{\it Correlation Function and Power Spectra in Cosmology},
Encyclopedia of Astronomy and Astrophysics, (IOP, London, UK, 2006);
ISBN 0333750888.

\bibitem{lon07}
M.~Longair, {\it Galaxy Formation}, (Springer Publishing, New York,
NY, 2007), 2nd ed..

\bibitem{teg02}
M.~Tegmark and M.~Zaldarriaga, 
{\it Separating the Early Universe from the Late Universe: cosmological 
parameter estimation beyond the black box}, 
Phys.\ Rev.\ {\bf D 66}, 103508 (2002).

\bibitem{teg04}
M.~Tegmark et al.,
{\it The 3D Power Spectrum of Galaxies from the SDSS},
Astrophys.\ J.\ {\bf 606}, 702-740 (2004) [astro-ph/0310725].

\bibitem{dur14}
A.~Durkalec et al.,
{\it The evolution of clustering length, large-scale bias and 
host halo mass at $2 < z < 5 $ in the VIMOS Ultra Deep Survey (VUDS)},
% gamma = 1.8 r0= 3.95 h^-1 \, Mpc
arXiv:1411.5688 [astro-ph.CO] (2014).

\bibitem{wan13}
Y.~Wang, R.~J.~Brunner and J.~C.~Dolence,
{\it The SDSS Galaxy Angular Two-Point Correlation Function},
Mon.\ Notice Astron.\ Soc.\ 432, 1961 (2013);
arXiv:1303.2432 [astro-ph.CO] (2013).

\bibitem{coi12}
A.~L.~Coil,
{\it Large Scale Structure in the Universe},
in {\sl Planets, Stars, and Stellar Systems}, edited by T.~D.~Oswalt
and W.~C. ~Keel (Springer, New York, USA), vol. 8;
arXiv:1202.6633 [astro-ph.CO] (2012).

% Original inflation papers

\bibitem{gut81}
A.~H.~Guth,
{\it Inflationary Universe: A possible Solution to the Horizon and Flatness Problems},
Phys.\ Rev.\ {\bf D23}, 347-356 (1981).

\bibitem{lin82}
A.~D.~Linde,
{\it A new Inflationary Universe Scenario: a Possible Solution of the Horizon, 
Flatness, Homogeneity, Isotropy, and Primordial Monopole Problems},
Phys.\ Lett.\ {\bf 108B}, 389 (1982).

\bibitem{alb82}
A.~Albrecht and P.~J. ~Steinhardt,
{\it Cosmology for Grand Unified Theories with Radiation Induced Symmetry Breaking},
Phys.\ Rev.\ Lett.\ {\bf 48}, 1220 (1982).

% More inflation references

\bibitem{lid00}
A.~R.~Liddle and D.~H.~Lyth,
{\it Cosmological Inflation and Large-Scale Structure},
(Cambridge University Press, 2000).


% QG on CMB - Hamber and Yu

\bibitem{hyu18}
H.~W.~Hamber and L.~H.~S.~Yu,
{\it Gravitational Fluctuations as an Alternative to Inflation},
Universe {\bf 2019}, 5(1), 31; arXiv:1807.10704v3 [gr-qc] (2018).

% Vacuum condensate picture

\bibitem{ham17}
H.~W.~Hamber,
{\it Vacuum Condensate Picture of Quantum Gravity},
Invited talk at the 2015 {\sl Coral Gables (Miami) International Conference on Particle Physics},
71 pp., arXiv [hep-th] 1707.08188 (2017), and references therein.
Published in {\it Symmetry and Quantum Gravity}, G.~Modanese ed., 
Symmetry {\bf 11(1)}, 87 (2019); DOI : 10.3390/sym11010087.

% Book 

\bibitem{book}
H.~W.~Hamber, {\sl Quantum Gravitation}, Springer Tracts in Modern Physics 
(Springer Publishing, Berlin, Germany and New York, NY, USA, 2009).

% Planck 2018 data

\bibitem{planck18}
Planck Collaboration: Y.~Akrami et~al.
{\it Planck 2018 results. I. Overview and the cosmological legacy of Planck},
arXiv:1807.06205v1 [astro-ph.CO] (2018).

% Hartree-Fock paper

\bibitem{hyu19hf}
H.~W.~Hamber and L.~H.~S.~Yu,
{\it Dyson's Equations for Quantum Gravity in the Hartree-Fock Approximation}, 
(2019), to appear.

% Feynman on gravity

\bibitem{fey63}
R.~P. Feynman.
{\it Quantum Theory of Gravitation},
Acta Phys.\ Polon.\  {\bf 24}, 697-722 (1963).

\bibitem{fey95}
R.~P.~Feynman, 
{\it Lectures on Gravitation}, Caltech lecture notes, 1962-1963;
{\sl edited by F.~B.~Morinigo, W.~G.~Wagner, and B.~Hatfield},
Advanced Book Program (Addison-Wesley, Boston, MA, USA, 1995).

\bibitem{dew62}
B.~S.~DeWitt,
{\it Quantization of Fields with Infinite-Dimensional Invariance Groups.
III. Generalized Schwinger-Feynman Theory},
J.\ Math.\ Phys.\ {\bf 3}, 1073 (1962).

\bibitem {lesh84}
H.~W.~Hamber, {\sl Simplicial Quantum Gravity},
in {\it Critical Phenomena, Random Systems and Gauge Theories}, 
{\sl 1984 Les Houches Summer School}, Session XLIII, 
edited by K.~Osterwalder and R.~Stora (North-Holland, Amsterdam, 1986).

% Wilson RG

\bibitem {wil72}
K.~G.~Wilson, 
{\sl Feynman-graph expansion for critical exponents},
Phys.\ Rev.\ Lett.\ {\bf 28}, 548 (1972);
{\sl Quantum field-theory models in less than 4 dimensions},
Phys.\ Rev.\ {\bf D 7}, 2911 (1973).

\bibitem{par73}
G.~Parisi, {\sl On the Renormalizability of not Renormalizable Theories},
Lett.\ Nuovo Cimento {\bf 6S2}, 450 (1973);
{\sl Theory of Non-Renormalizable Interactions - The large N Expansion},
Nucl.\ Phys.\ {\bf B 100}, 368 (1975);
{\sl Symanzik's Improvement Program},
Nucl.\ Phys.\ {\bf B 254}, 58 (1985).

\bibitem {par76}
G.~Parisi, {\sl On Non-Renormalizable Interactions}, 
Proceedings of the 1976 Carg\'ese NATO Advances Study Institute,
on {\it New Developments in Quantum Field Theory and Statistical Mechanics} 
edited by M.~Levy and P.~Mitter (Plenum Press, New York, 1977).

% General Statistical Field Theory References

\bibitem {par81}
G.~Parisi, {\sl Statistical Field Theory}; 
(Benjamin Cummings, San Francisco, CA, USA, 1981).

\bibitem {itz91}
C.~Itzykson and J.~M.~Drouffe, {\sl Statistical Field Theory},
(Cambridge University Press, Cambridge, UK, 1991).

\bibitem {car96}
J.~L.~Cardy, {\sl Scaling and Renormalization in Statistical Physics},
Cambridge Lecture Notes in Physics, (Cambridge University Press, Cambridge, UK, 1996).

% Zinn-Justin book

\bibitem {zin02}
J.~Zinn-Justin, {\sl Quantum Field Theory and Critical Phenomena}, 4th ed.; 
Oxford University Press (Oxford, UK, 2002), and further references therein.

\bibitem {bre10}
E.~Brezin, {\sl Introduction to Statistical Field Theory}, 
(Cambridge University Press, Cambridge, UK, 2010).

% Invariant R-R Correlations in Quantum Gravity

\bibitem {cor94}
H.~W.~Hamber,
{\it Invariant Correlations in Simplicial Gravity},
Phys.\ Rev. {\bf D50}, 3932 (1994).

% Our G(box) cosmology work

\bibitem{hw05}
H.~W.~Hamber and R.~M.~Williams,
{\it Nonlocal effective gravitational field equations and the running of Newton's G},
Phys.\ Rev.\ {\bf D72}, 044026 (2005).

\bibitem{rei10}
H.~W. ~Hamber and R.~Toriumi, 
{\it Cosmological Density Perturbations with a Scale-Dependent Newton's G}, 
Phys.\ Rev.\ {\bf D82}, 043518, 2010; 
{\it Scale-Dependent Newton's Constant G in the Conformal Newtonian Gauge},
Phys.\ Rev.\ {\bf D84}, 103507, 2011.

% More G(box) work,  specifically the nonperturbative scale xi

\bibitem {rei14}
H.~W.~Hamber and R.~Toriumi,
{\it Inconsistencies from a Running Cosmological Constant},
36 pp., AEI preprint January 2013. Review article 
published in the Int.\ J.\ Mod.\ Phys.\ {\bf D 22} (2013);
% On the Cosmological Constant: its Identification as a Renormalization
% Group Invariant Scale Corresponding to a Gravitational Condensate
{\sl Frontiers of Fundamental Physics}, {\it PoS}, FFP14:178 (2016).

% Scaling Exponents for Lattice Gravity in 4D

\bibitem {ham15}
H.~W.~Hamber,
{\it Scaling Exponents for Lattice Quantum Gravity in Four Dimensions},
Phys.\ Rev.\ {\bf D 92}, 064017 (2015); see also 
{\it Phases of Simplicial Quantum Gravity in Four Dimensions: Estimates for the Critical Exponents},
Nucl.\ Phys. {\bf B400}, 347 (1993), and 
{\it Gravitational Scaling Dimensions},
Phys.\ Rev.\ {\bf D 61}, 124008 (2000).

% Relationship between average Curvature, Lambda and xi

\bibitem {loops}
H.~W.~Hamber and R.~M.~Williams,
{\it Gravitational Wilson Loop and Large Scale Curvature},
Phys.\ Rev.\ {\bf D 76} 084008 (2007);
{\it Gravitational Wilson Loop in Discrete Gravity},
Phys.\ Rev.\ {\bf D 81} 084048 (2010).

% 2+epsilon Expansion

\bibitem{wei79}
S.~Weinberg,
{\sl  Ultraviolet Divergences in Quantum Gravity},
in  'General Relativity - An Einstein Centenary Survey', edited by S.~W.~Hawking and W.~Israel,
(Cambridge University Press, Cambridge, UK, 1979).

\bibitem {gas78}
R.~Gastmans, R.~Kallosh and C.~Truffin,  
{\it Quantum Gravity Near Two Dimensions},
Nucl.\ Phys.\ {\bf B 133} 417
(1978); \\
S.~M.~Christensen and M.~J.~Duff,  
{\it Quantum Gravity in $2+\epsilon$ Dimensions},
Phys.\ Lett.\ {\bf B 79} (1978) 213.

% 2+eps expansion of nu to one and two loops

\bibitem {eps}
H.~Kawai and M.~Ninomiya, 
{\it Renormalization Group and Quantum Gravity},
Nucl.\ Phys.\ {\bf B336}, 115 (1990); \\
H.~Kawai, Y.~Kitazawa and M.~Ninomiya, 
{\it Scaling Exponents in Quantum Gravity near two Dimensions},
Nucl.\ Phys.\ {\bf B393}, 280 (1993),
and 
{\it Ultraviolet Stable Fixed Point and Scaling Relations in
$(2+\epsilon)$-dimensional Quantum Gravity},
{\bf B404} 684 (1993); \\
Y.~Kitazawa and M.~Ninomiya, 
{\it Scaling Behavior of Ricci Curvature near two Dimensions},
Phys.\ Rev.\ {\bf D55}, 2076 (1997); \\
T.~Aida and Y.~Kitazawa, 
{\it Two Loop Prediction for Scaling Exponents in $(2+\epsilon)$-dimensional 
Quantum Gravity},
Nucl.\ Phys.\ {\bf B491}, 427 (1997).

% Large d result 1/nu = d-1 

\bibitem {larged}
H.~W.~Hamber and R.~M.~Williams, 
{\it Non-Perturbative Gravity and the Spin of the Lattice Graviton},
Phys.\ Rev.\ {\bf D 70}, 124007 (2004);
{\it Quantum Gravity in Large Dimensions},
Phys.\ Rev. \ {\bf D 73}, 044031 (2006).

% Exact value for nu in 2+1 dimensions

\bibitem {htw12}
H.~W.~Hamber, R.~Toriumi, and R.~M.~Williams,
{\it Wheeler-DeWitt Equation in 2+1 Dimensions},
Phys.\ Rev.\  {\bf D 86}, 084010 (2012);
{\it Wheeler-DeWitt Equation in 3+1 Dimensions},
Phys.\ Rev.\  {\bf D 88}, 084012 (2013).

% E-H truncation references

\bibitem{reu98}
M.~Reuter,
{\it Nonperturbative Evolution Equation for Quantum Gravity},
Phys.\ Rev.\  {\bf D 57}, 971 (1998); \\
% \bibitem{reu08}
M.~Reuter and H.~Weyer,
{\it The Role of Background Independence for Asymptotic 
Safety in Quantum Einstein Gravity},
% in {\sl Quantum Gravity: Challenges and Perspectives},
% Bad Honneff (Hermann Nicolai ed.), 
General Relativ.\ Gravit.\  {\bf 41}, 983 (2009),
and further references therein.

\bibitem{lit04}
D.~F.~Litim, 
{\it Fixed Points of Quantum Gravity},
Phys.\ Rev.\ Lett.\ {\bf 92} 201301 (2004); \\
P.~Fischer and D.~F.~Litim,
{\it Fixed Points of Quantum Gravity in Extra Dimensions},
Phys.\ Lett.\  {\bf B 638}, 497 (2006).

% More Recent EH Truncation

\bibitem{reu14} 
D.~Becker and M.~Reuter, 
{\it En Route to Background Independence: Broken Split-Symmetry, 
and how to Restore it with Bi-Metric Average Actions},
Annals Phys.\  {\bf 350}, (2014) 225.

\bibitem{fal15}
K.~Falls, 
% ``On the Renormalization of Newton's constant''
%arXiv:1410.4815 [hep-th];
{\it Critical scaling in Quantum Gravity from the Renormalisation Group},
arXiv:1503.06233 [hep-th]; \\
K.~Falls, D.~F.~Litim, K.~Nikolakopoulo and C.~Rahmede,
{\it Further Evidence for Asymptotic Safety of Quantum Gravity},
arXiv:1501.05331 [hep-th].

\bibitem{per16}
N.~Ohta, R.~Percacci and A.~D.~Pereira,
{\it Gauges and Functional Measures in Quantum Gravity I: Einstein Theory},
JHEP {\bf 1606}, 115 (2016).

\bibitem{gie15} 
H.~Gies, B.~Knorr and S.~Lippoldt,
{\it Generalized Parametrization Dependence in Quantum Gravity},
Phys.\ Rev.\ D {\bf 92}, no. 8, 084020 (2015).	

% Weinberg cosmology book

\bibitem{wei08}
S.~Weinberg,
{\it Cosmology}, (Oxford University Press, Oxford, UK, 2008).

% Dodelson cosmology book

\bibitem{dod03}
S.~Dodelson,
{\it Modern Cosmology}, (Academic Press, Amsterdam, The Netherlands, 2003).

% Index ns predictions from inflation

\bibitem{ste04}
P.~J.~Steinhardt,
{\it Cosmological Perturbations},
Mod.\ Phys.\ Lett.\ A {\bf 19}, 967-982, 2004.

% Recent SDSS data 

\bibitem{gil18}
H.~Gil-Marín et~al,
{\it The Clustering of the SDSS-IV Extended Baryon Oscillation
Spectroscopic Survey DR14 Quasar Sample: Structure Growth Rate Measurement
from the Anisotropic Quasar Power Spectrum in the Redshift Range $0.8<z<2.2$},
SDSS collaboration, (2018);
Mon.\ Notices R.\ Astron.\ Soc.\ {\bf 437}, 4773-4794 (2017).
% First author on the above "gil18" ref. is actually M. Ata, F. Baumgarten, ...

% Transfer function T(k)

\bibitem{whu97}
W.~Hu and D.~J.~Eisenstein,
{\it Baryonic Features in the Matter Power Spectrum},
ArXiv astro-ph/9709112;
\url{http://background.uchicago.edu/~whu/transfer/transferpage.html}
(accessed on 6-15-2019).

\bibitem{hz70}
E.~R.~Harrison, 
{\it Fluctuations at the Threshold of Classical Cosmology},
Phys.\ Rev.\ {\bf D 1}, 2726 (1970); \\
Y.~B.~Zel'dovich,  
{\it A Hypothesis Unifying the Structure and the Entropy of the Universe},
Mon.\ Not.\ Roy.\ Astron.\ Soc.\ {\bf 160}, p. 1P (1972); \\
P.~J.~E.~Peebles and J.~T.~Yu, 
{\it Primeval Adiabatic Perturbation in an Expanding Universe},
Astrophys.\ J.\  {\bf 162}, 815 (1970).

\bibitem{ste06}
L.~A. ~Boyle, P.~J. ~Steinhardt, and N.~Turok,
{\it Inflationary Predictions for Scalar and Tensor Fluctuations Reconsidered},
Phys.\ Rev.\ Lett.\  {\bf 96} (11):111301 (2006).

% One more inflation paper 

\bibitem{gut14}
A.~H.~Guth, D.~I.~Kaiser, and Y.~Nomura,
{\it Inflationary Paradigm after Planck 2013},
Phys. \ Lett. \ {\bf B}, 733:112-119 (2014).

% inflation debate

\bibitem{ste12}
P.~J.~Steinhardt,
{\it The Cyclic Theory of the Universe}, Princeton University Press, NJ, USA (2012); \\
P.~J.~ Steinhardt and N.~Turok, 
{\it A Cyclic Model of the Universe}
Science {\bf 296}, 1436-1439 (2002).

\bibitem{teg05}
M.~Tegmark,
{\it What does Inflation Really Predict?},
J.\ Cosmol.\  Astropart.\ Phys.\ (JCAP) {\bf 4} 001 (2005).

\bibitem{ste13}
A.~Ijjas, P.~J.~Steinhardt, and A.~Loeb,
{\it Inflationary Paradigm in Trouble after Planck 2013},
Phys.\ Lett.\ {\bf B} 723, 261-266 (2013).

\bibitem{ste14}
A.~Ijjas, P.~J.~Steinhardt, and A.~Loeb,
{\it Inflationary Schism},
Phys. \ Lett.\ {\bf B} 736, 142-146, (2014).

\bibitem{ste17}
P.~J.~Steinhardt, A.~Ijjas and A.~Loeb,
{\it Pop Goes the Universe},
Sci.\ Am.\ {\bf 316}, 32-39 (2017).

% Early Universe - Veneziano

\bibitem{ven02}
G.~Veneziano,
{\it String Cosmology: The Pre-Big Bang Scenario}, 
CERN-Th/2000-42, hep-th/0002094; published in 
{\sl The Primordial Universe}, Les Houches Summer School LXXI, 1999,
P.~Binetruy, R.~Schaefer, J.~Silk and F.~David editors
(Springer Verlag, Berlin/Heidelberg, Germany, 2000).

\bibitem{ven04}
G.~Veneziano, 
{\it The Myth of the Beginning of Time}, Sci. Am. {\bf 290}, 54-65, (2004). 

\bibitem{ven07} 
M.~Gasperini and G.~Veneziano,
{\it Pre-Big Bang in String Cosmology},
hep-th/9211021; Astropart.\ Phys.\ 1, 317, (1993);
% ``Pre-Big Bang Scenario in String Cosmology,''
hep-th/0703055;  \\
A.~Buonanno, T.~Damour and G.~Veneziano,
{\it Pre-Big Bang Bubbles from the Gravitational Instability of Generic String Vacua},
CERN-Th/98-187; Nucl.\ Phys.\ {\bf B543} (1999) 275-320.

% Early Universe - Hartle

\bibitem{har83}
J.~B.~Hartle and S.~W.~Hawking,
{\it Wave function of the Universe},
Phys.\ Rev.\ {\bf D 28} 2960 (1983).

\bibitem{har08}
J.~B.~Hartle, S.~W.~Hawking and T.~Hertog,
{\it No-Boundary Measure of the Universe},
Phys.\ Rev.\ Lett. {\bf 100}, 201301 (2008).

% More Alternatives to inflation

\bibitem{hol02}
S.~Hollands and R.~M.~Wald,
{\it An Alternative to Inflation},
ArXiv: gr-qc/0205058 (2002).

% Quantum mechanical origin of cosmological  fluctuations in matter & CMB

\bibitem{muk81}
S.~V.~Mukhanov and G.~V.~Chibisov, 
{\it Quantum Fluctuations and a Nonsingular Universe},
Sov.\ Phys.\ JETP Lett. {\bf 33}, 532 (1981); \\
S. ~Hawking, 
{\it The Development of Irregularities in a Single Bubble Inflationary Universe},
Phys.\ Lett. {\bf 115B}, 295 (1982); \\
A.~A.~Starobinsky, 
{\it Dynamics of Phase Transition in the New Inflationary Universe 
Scenario and Generation of Perturbations},
Phys.\ Lett. {\bf 117B}, 175 (1982); \\
A.~Guth and S~-Y.~Pi, 
{\it Fluctuations in the New Inflationary Universe},
Phys.\ Rev.\ Lett. {\bf 49}, 1110 (1982); \\
J.~M.~Bardeen, P.~J.~Steinhardt, and M.~S.~Turner, 
{\it Spontaneous Creation of Almost Scale-free Density 
Perturbations in an Inflationary Universe},
Phys.\ Rev.\ D {\bf 28}, 679 (1983); \\
W.~Fischler, B. ~Ratra, and L.~Susskind, 
{\it Quantum Mechanics of Inflation},
Nucl.\ Phys.\ B {\bf 259}, 730 (1985).


%%%%%% Outdated Planck data, not used here %%%%%%%% 

% P(k) plot with Planck 2015 data

%\bibitem{mac15}
%N.~MacCrann et~al.,
%{\it Cosmic Discordance: Are Planck CMB and CFHTLenS weak lensing 
%measurements out of tune?},
%arXiv:1408.4742v2 [astro-ph.CO] (2015).

%%%%%% Reference From previous paper, not used here %%%%%%%% 

% Damour review

%\bibitem{dam06}
%T.~Damour, 
%{\it Experimental Tests of Gravitational Theory},
%in Review of Particle Physics, J.\ Phys.\ {\bf G} 33, 1 (2006);
%update in \\
%\url{http://pdg.lbl.gov/2009/reviews/rpp2009-rev-gravity-tests.pdf} (Nov. 2009).

% Renormalons 

%\bibitem{ben99}
%M.~Beneke, {\sl Renormalons}, CERN-TH/98-233, Physics Reports {\bf 317}, 1 (1999), and references therein.


\end{thebibliography}
\end{document}